\documentclass[review]{elsarticle}

\usepackage[maxfloats=25]{morefloats}
\usepackage{lineno,hyperref}
\usepackage{graphicx}
\usepackage{amssymb}
\usepackage{amsmath}
\usepackage{bm}
\usepackage[usenames, dvipsnames]{color}
\usepackage[scale=0.75,twoside,bindingoffset=5mm]{geometry}
\modulolinenumbers[5]

\usepackage{changes}

\journal{Journal of \LaTeX\ Templates}

\begin{document}

\begin{frontmatter}

\title{Nonlinear, Nonequilibrium and Collective Dynamics in a Periodically Modulated Cold Atom System}
%\tnotetext[mytitlenote]{Fully documented templates are available in the elsarticle package on \href{http://www.ctan.org/tex-archive/macros/latex/contrib/elsarticle}{CTAN}.}

\author[cor]{Geol Moon}
\author[Myoung-Sun Heo]{Myoung-Sun Heo\corref{cor1}}
\ead{hms1005@kriss.re.kr}
\author[Yonghee Kim]{Yonghee Kim}
\author[Heung-Ryoul Noh]{Heung-Ryoul Noh}
\author[cor]{Wonho Jhe\corref{cor1}}
\ead{whjhe@snu.ac.kr}

\cortext[cor1] {Corresponding author}
\address[cor]{Department of Physics and Astronomy, Institute of Applied Physics, Seoul National University, Seoul 08826, Korea}
\address[Myoung-Sun Heo]{Korea Research Institute of Standards and Science, Daejeon 34113, Korea}
\address[Yonghee Kim]{Quantum Optics Division, Korea Atomic Energy Research Institute, Daejeon 34057, Korea}
\address[Heung-Ryoul Noh]{Department of Physics, Chonnam National University, Gwangju 61186, Korea}

\begin{abstract}
The physics of critical phenomena in a many-body system far from thermal equilibrium is an interesting and important issue to be addressed both experimentally and theoretically.   
The trapped cold atoms have been actively used as a clean and versatile simulator for classical and quantum-mechanical systems, deepening understanding of the many-body physics behind. Here we review the nonlinear and collective dynamics in a periodically modulated magneto-optically trapped cold atoms. By temporally modulating the intensity of the trapping lasers with the controlled phases,  one can realize two kinds of nonlinear oscillators, the parametrically driven oscillator and the resonantly driven Duffing oscillator, which exhibit the dynamical bistable states. Cold atoms behave not only as the single-particle nonlinear oscillators, but also as the coupled oscillators by the light-induced inter-atomic interaction, which leads to the phase transitions far from equilibrium in a way similar to the phase transition in equilibrium. The parametrically driven cold atoms show the ideal mean-field symmetry-breaking transition, and the symmetry is broken with respect to time translation by the modulation period. Such a phase transition results from the cooperation and competition between the inter-particle interaction and the fluctuations, which lead to the nonlinear switching of atoms between the vibrational states, and the experimentally measured critical characteristics prove it as the ideal mean-field transition class. On the other hand, the resonantly driven cold atoms that possess the coexisting periodic attractors exhibit the kinetic phase transition analogous to the discontinuous gas-liquid phase transition in equilibrium, and interestingly the global interaction between atoms causes the shift of the phase-transition boundary. We demonstrate that the temporally driven cold atom system serves as a unique and controllable platform suitable for investigating the nonlinear dynamics of many-body cold atoms far from equilibrium and for relating the relevant dynamics to other domains of physics. The results presented in this article may be useful for better understanding of the fundamentals of critical phenomena occurring in a many-body system far from thermal equilibrium, which still demands further studies both experimental and theoretical. 
\end{abstract}


\end{frontmatter}

\tableofcontents
%\linenumbers

\section{Introduction}
\label{Introduction}

Thermal and statistical physics have been successful for describing the classical and quantum systems in thermal equilibrium~\cite{R. Kubo 1995, Katja Lindenberg 1990, Landau:StatMech}. However, most of processes that occur in nature are not in equilibrium. For example, numerous macroscopic biological systems such as the birds, fish and insects can survive against the upper-level predators by driving themselves out of equilibrium or displaying the collective behavior called \textit{flocking}~\cite{Philip Ball 2004, Arkady Pikovsky 2001, Hermann Haken 1983}.  In addition, any transient systems before reaching their equilibrium or the turbulent systems consist of other nonequilibrium classes~\cite{Pottier 2010, Frank Moss 1989, Jean-Louis Barrat 2003, Stephen B. Pope 2000, Luca Cipellette 2002}.
With the recent advancement of the experimental capabilities to look into the smaller systems at the higher spatial and temporal resolutions with a better controllability, one can probe the nonequilibirum features of the dynamics systems, and observe their similarities to those in thermal equilibrium or find the unique features such as the directed transports~\cite{Renzoni2009} and the discrete time crystals~\cite{,Khemani2016, Zhang2016,Choi2017}. The results have led one to develop new concepts for the microscopic nonequilibrium statistical physics including the nonequilibrium steady-states~\cite{Speck2006}, the fluctuation theorem~\cite{Gallavotti1995, Evans2002,CarlosBustamanteandFelixRitort2005}, the Jarzynski’s equality~\cite{Jarzynski1997} and the Crooks equation~\cite{Crooks1999}.%, and provided us with thermodynamics correspondence between classical and quantum mechanical phenomena ~\cite{Eisert2015,Jarzynski2015}. 

To investigate experimentally the nonequilibrium phenomena, researchers have used various kinds of small systems, ranging from the biological ones such as the RNA or DNA, and the protein molecules \cite{Liphardt2002,Junier2009,Collin2005,Ioan 2004} to the physical ones including the cold atoms, ions and electrons~\cite{Renzoni2009,Zhang2016,M. Grifoni,L J Lapidus PRL 1999,Cross rev 1993,Dehmelt rev 1990,Gabrielse PRL 1985,Creffield2009,DeChiara2015,An2014}, the colloidal particles~\cite{victor,Wang2002}, the mechanical oscillators~\cite{Chan PRL 2008,Chan PRL 2007,Chan PRL 2006,Chan PRB 2006,J S Aldridge PRL 2005,R L Badzey Appl Phys Lett 2004,Douarche2005}, the electronic circuits~\cite{Dykman PRL 1990,Garnier2005,Pekola2015,Saira2012}, the diamond defect centers~\cite{Schuler2005} and the optomechanical systems~\cite{Aspelmeyer2014}. Among them, the trapped cold atoms and ions have recently served as  the versatile and clean platforms to simulate and investigate statistical physics~\cite{Renzoni2009,Zhang2016,Creffield2009,DeChiara2015,An2014,Dawkins2016,Brantut2013} as well as condensed matter physics ~\cite{Esslinger2010,Bloch2008,Dalibard2011} by virtue of the unprecedented precise controllability of the internal as well as the external degrees of freedom.

In this review, we discuss the collective and nonequilibrium phenomena that occur in the periodically driven cold atom system with the specific interparticle interaction and the thermal fluctuations. To achieve this, we first realize the magneto-optical atom trap  as a many-body interacting system consisting of $10^6\sim10^8$ $^{85}$Rb (Rubidium) neutral atoms. The temperature of trapped atoms is in the range of a few hundreds $\mu$K, and the intensity of trapping lasers is modulated periodically for the realization of the driven nonlinear oscillators such as the parametrically driven as well as the resonantly driven oscillator. One can deal with the time evolution of the system using the fluctuation-induced switching of particles between the dynamics attractors caused by the thermal fluctuations~\cite{JETF 1979,Kunz,Horstheemke,Dykman PRE 1994,Drummond1980,Lugiato1984,moss vol2}. Although the corresponding theory looks similar to the classical reaction rate theory~\cite{Hanggi1990} represented by the Kramers' seminal work,  one still needs completely new approaches because the systems of our interest are not in thermal equilibrium. The novel methods have been indeed employed to investigate the driven nonlinear oscillators in the presence of fluctuations such as the Josephson junctions, the nano-magnetic oscillators and the microwave cavities~\cite{Dykman2012}. We have analyzed the driven cold atoms using the similar approaches and we have observed,  in particular, the unique collective phenomena that result from the competitive and cooperative interplays between the fluctuations and the atom-atom interactions~\cite{K Kim PRL 2006,M Heo PRE 2010,gmoon njp2013}. 

This paper is organized as follows. In Sec.~2, we describe the driven nonlinear dynamics realized in two types of modulated cold atoms; the parametrically driven Duffing oscillator (PDDO) and the resonantly driven Duffing oscillator (RDDO). Such nonlinear dynamic systems possess  two bistable states, but the characteristics of the bistable states in each nonlinear system are quite different from each other, bringing in the distinct collective responses to the atom-atom interactions. In Sec.~3, we provide the general theoretical tools to understand the fluctuation-induced switching between the bistable states, which includes calculation of the most probable switching paths~\cite{Chan PRL 2008,JETF 1979,J. Hales 2000,M. I. Dykman 5902} and the corresponding activation barriers. The latter part of this section deals with how one modifies the switching dynamics that result from the long-range atom-atom interaction. Sections 4 and 5 discuss the detailed results of the atom-atom interaction for the PDDO and the RDDO, respectively.  We discuss the spontaneous breaking of the discrete time-translation symmetry in the PDDO and compare it with the critical phenomena in thermal equilibrium systems. In Sec.~5, using the experimentally realized RDDO, we describe the asymmetry in the effects of the interaction due to the differences in the switching paths, which results in the kinetic phase transition analogous to the gas-liquid phase transition in thermal equilibrium.  The summary and outlook in the final section closes the review.

\section{Nonlinear dynamics in a modulated cold atom system}
\label{sec2}

We consider the cold atomic system which consists of the conventional six-beam magneto-optical trap (MOT). The atomic motion inside the MOT is described by the interaction between atoms and photons, called the radiation (scattering) force. When an atom absorbs photons of frequency $\omega_L$, each absorbed photon transfers its momentum $\hbar k$ to the atom ($k$ is the photon wave vector, $k=\omega_L/c$). The radiation force on the two-level atom due to the corresponding momentum kicks is given by the relation, (photon momentum) $\times$ (photon scattering rate), 
\begin{equation}
	 	F_{scatt} =\hbar k \times\frac{\Gamma}{2}\frac{s_0 }{1+s_0+(2 \Delta / \Gamma)^2},
\end{equation}
where $\Gamma$ is the natural linewidh or the spontaneous decay rate of the atomic excited state, $\Delta=\omega_L-\omega_A$ is the frequency detuning of the laser relative to the atomic resonance frequency $\omega_A$, and $s_0$ is the laser intensity normalized to the saturation intensity $I_s$ of the two-level atom~\cite{Metcalf,AtomicPhysics,coldAM}. The MOT consists of three pairs of counterpropagating laser beams and  an anti-Helmholtz coil providing the  magnetic field gradient $b$. We only consider the one-dimensional atomic motion along the axis of the anti-Helmholtz coil, $z$-axis (refer to Fig. 1) for simplicity (we indeed modulate parametrically the intensities of only one pair of counterpropagating trap lasers). Notice that the trap laser intensities are weak enough that the Rabi frequency is smaller than the natural linewidth and thus the two-level approximation is valid. The force $F_{\pm}$ on the atoms by the laser beams propagating along the $\pm z$ direction is given by~\cite{Metcalf,AtomicPhysics,coldAM},
\begin{equation}
F_{\pm} =\pm\frac{\hbar k\Gamma}{2}\frac{s_0}{1+s_0+(2\Delta_{\pm} / \Gamma)^2},
\end{equation}
where the frequency detunings $\Delta_{\pm}$ for each counterpropagating beams are,
\begin{equation}
\Delta_{\pm}=\Delta \mp k \dot z \mp \frac{\mu_B b}{\hbar} z.
\end{equation}
Here $\mu_B$ is the Bohr magneton and $b$ is the magnetic field gradient. Note that the $\pm k \dot z$ term represents the frequency shift due to the Doppler effect and $\pm \frac{\mu_B b}{\hbar} z$ describes the Zeeman shift of the atomic levels. Then, the equation of  motion of the atom irradiated by two counterpropagating lasers becomes,
\begin{equation}
\label{MOTeq} m_a\ddot{z}=F_++F_-=\frac{\hbar k \Gamma}{2}\left[\frac{s_0}{1+s_0+\frac{4}{\Gamma^2}(\Delta-k \dot{z}-\frac{\mu_{B}b}{\hbar} z)^2}
-\frac{s_0}{1+s_0 +\frac{4}{\Gamma^2}(\Delta+k \dot{z}+\frac{\mu_{B}b}{\hbar} z)^2}\right],
\end{equation}
where $m_a$ is the atomic mass.
 
 When Eq. (\ref{MOTeq}) is expanded in a power series of $z$ and $\dot z$ in the limit of small values of $s_0$, one can obtain the following approximate equation,
\begin{equation}
\label{nonlinearwoforce}\ddot{z}+\gamma\dot{z}+\omega_0^2 z +B_0\left(z+\frac{\hbar k}{\mu_{B}b}\dot{z}\right)^3 =0,
\end{equation}
where $\omega_0$ is the trap frequency, $\gamma$  the damping coefficient and $B_0$  the nonlinear coefficient. These trap parameters are expressed by,
\begin{eqnarray}
\label{parameter}
\omega_0&=&\sqrt{\frac{8k\mu_B b
s_0(-\Delta/\Gamma)}{m_a (1+4(\Delta/\Gamma)^2)^2}}, \\
\gamma&=&\frac{\hbar k}{\mu_B b}\omega_0^2,\\
B_0&=&\frac{8\mu_{B}^{2}b^2(4(\Delta/\Gamma)^2-1)}{\hbar^{2}\Gamma^{2}(4(\Delta/\Gamma)^2+1)^{2}}\omega_{0}^{2}.
\end{eqnarray}
  According to Eq.~(\ref{nonlinearwoforce}), the atomic motion can be described by the damped harmonic oscillator with a cubic nonlinearity. Therefore one can study the nonlinear dynamics in this cold atomic system by periodically modulating the parameters such as the laser intensities, the magnetic field gradient and the detuning of trap lasers.

 The cold atom system has several advantages to study the nonlinear dynamics compared to other systems such as the MEMS, analog circuit and biological system. For example, the trap parameters (e.g., the damping coefficient, trap frequency and nonlinearity coefficient) are not fixed, and thus one can change these parameters over a broad range. This means that one can explore the large parameter space, which allows the experimental disclosure of the various nonlinear dynamic characteristics. One can also realize different types of  nonlinear systems, the PDDO and the RDDO, by adjusting the frequency and  phase of external modulation, as described in Sec.~\ref{sec2_setup} and Sec.~\ref{sec2_nonlineardynamics}. Finally, since this system consists of  about $10^7$ atoms, one can obtain the ensemble properties of the atomic motion instead of the single particle trajectories. In this section, we briefly introduce the experimental setup that allows investigation of various nonlinear dynamics that occur in the driven cold atom system.  
 
\subsection{Experimental setup}
\label{sec2_setup}

\begin{figure}[thb] \centering
\includegraphics[scale=0.35]{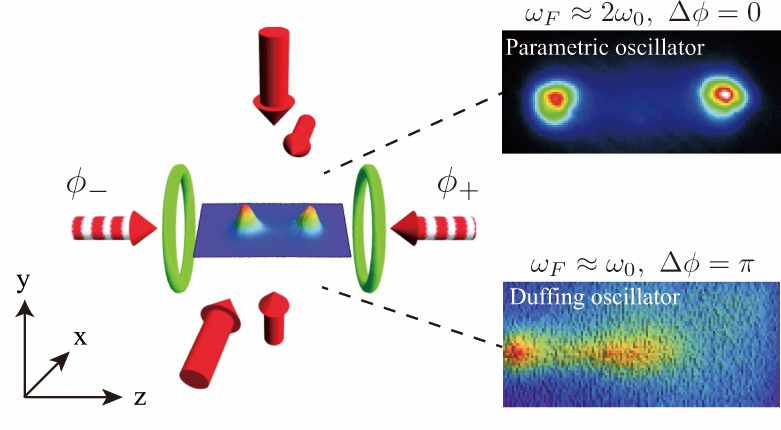}
\caption[figure1]{\label{figure1}\small Experimental setup for the conventional six-beam magneto-optical trap (MOT): The intensity of trapping lasers counterpropagating along the $z$-axis is periodically driven at the modulation frequency $\omega_F$ with the phase difference of $\Delta\phi=\phi_+-\phi_-$. The atomic cloud is resonantly excited to two kinds of nonlinear oscillators, the PDDO and the RDDO, depending on  $\omega_F$ and $\Delta\phi$.}
\end{figure}

 To study the nonlinear dynamics and nonequilibrium critical phenomena using cold atoms, we used the conventional six-beam MOT of $^{85}$Rb atoms as shown in Fig.~\ref{figure1}. We confine the center of atoms to move dominantly along the $z$-axis by setting up the trap laser intensities along the $z$-axis (longitudinal axis) 5 times weaker than those along the other axes (transverse axes) to restrict the atomic motion virtually to one dimension. We  also set the frequency detuning $\Delta$ of the trap laser along each axis different from one another to avoid the sub-Doppler trap~\cite{KihwanPRA2004,Heo2007,Noh2007}.
 
 Then we periodically modulated the trap parameters and observed the atomic motion $in situ$ using a fast charge-coupled-device (CCD) camera. For the trap parameters that depend on the laser intensity, magnetic field gradient and laser detuning, we study the atomic dynamics under the parametric excitation by numerical calculation as well as experimental measurement. First, for the laser-intensity modulation, we periodically modulated independently each of the longitudinal trap-laser intensities  using acousto-optic modulators. Depending on the modulation frequency $\omega_F$ and the relative phase difference of the counterpropagating laser beams $\Delta\phi$, we observe the PDDO ($\omega_F\approx 2\omega_0$ and $\Delta\phi=0$)~\cite{K Kim PRL 2006,M Heo PRE 2010,Nohjkps2005} and the RDDO ($\omega_F\approx\omega_0$ and $\Delta\phi=\pi$)~\cite{gmoon njp2013,gmoon jkps2011} which is described in Sec.~\ref{sec2_nonlineardynamics}. For the magnetic field-gradient modulation, on the other hand, we constructed additional Helmholtz coil along the same $z$-axis as the MOT and applied the periodic driving current. For the modulation frequency $\omega_F$ and the modulation amplitude $\epsilon$, we studied the complex nonlinear dynamics such as the bifurcation map and the strange attractors. In particular,fFor the laser-detuning modulation, one can easily achieve such a scheme by applying the modulation signal to the laser current feedback, which exhibits the dynamics similar to the kicked rotor as described in Sec.~\ref{magnetic modulation}.
 
 In these nonlinear systems, the collective behaviors and critical phenomena occur because of the interplay between thermal fluctuations and atom-atom interactions. The thermal fluctuations come from  the random momentum kicks due to spontaneous emission of photons. To control the atomic interactions, we varied the total number of atoms by adjusting the intensity of the hyperfine repumping laser. 
 The density distribution of atoms was measured $in situ$ using a fast CCD camera (Sec.~\ref{sec3} and Sec.\ref{sec4}).

\begin{figure}[ht] \centering
\includegraphics[scale=0.27]{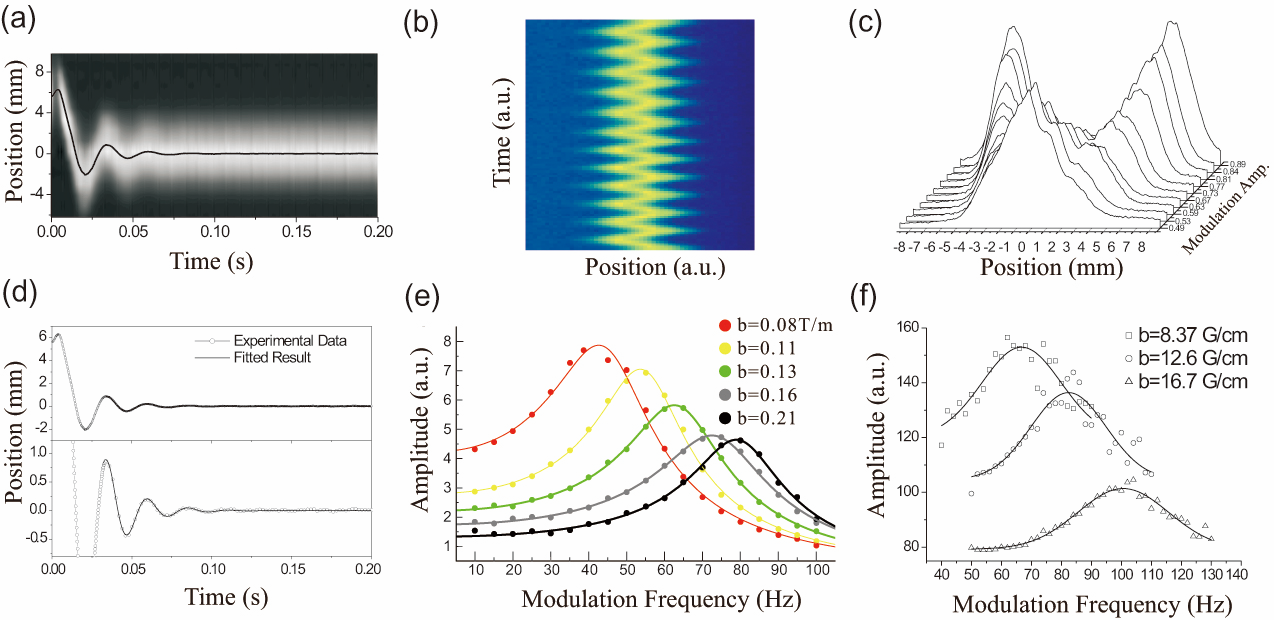}
\caption[figure2]{\label{fig:parametermeasure}\small Several schemes to measure the trap parameters by (a) transient oscillation, (b) forced harmonic oscillation and (c) parametric resonance. The panels in (a), (b), and (c) display the atomic motion obtained by each method. (d), (e), and (f) represent the theoretical fitting results with the experimental data, from which the trap parameters are estimated. The lower panel in (d) is the detailed plot after one period of the upper panel in (d). And in the cases of (e) and (f), the measurement variable  is the magnetic gradient field $b$. (Figures from Refs.~\cite{K. Kim 053406,K. Kim 003413,geolmoonpra2010})}
\end{figure}

 To describe the many-body dynamics in the MOT, it is essential to know the physical quantities such as the trap frequency, damping rate, lifetime and temperature. On the other hand, once one fully understands the many-body behaviors, one can extract the trap parameters by observing the atomic motion. We have measured the trap parameters, the trap frequency and the damping rate, in various ways such as the transient oscillation method~\cite{X. Xu 2002,K. Kim 053406,A. M. Steane 1991}, the parametric resonance method~\cite{K. Kim 003413} and the forced harmonic oscillation using the modulated magnetic-field and laser intensity~\cite{P. Kohns 1993,geolmoonpra2010} as shown in Fig.~\ref{fig:parametermeasure}.
 
Figures~\ref{fig:parametermeasure} (a), (b), and (c)  display, respectively, the detected atomic cloud motion in the transient oscillation method, the forced harmonic oscillation method and the parametric resonance method, while Figs.~\ref{fig:parametermeasure} (d), (e), and (f) show, respectively, the measurement of the vibrational amplitude versus the time and the driving frequency.
The trap frequency $\omega_0$ and the damping coefficient $\gamma$ are extracted by fitting the data to the theoretical results. The magneto-optically trapped neutral atoms are captured at the trap center where the magnetic field is zero. It means that if we shift the trap center by applying the uniform magnetic field  $B_z$ as the bias field, the atomic cloud also shifts its position along the direction of the bias field. When a uniform magnetic field is applied to the MOT, the position of the trap center is shifted by $B_z/b$, where $b$ is the magnetic field gradient in the $z$-axis of the MOT. As the uniform magnetic field is suddenly turned off, the atomic cloud returns to the original trap center. The above procedure describes the transient oscillation method, and since the trajectory of the atomic cloud usually undergoes the underdamped harmonic motion, the trajectory is given by,
\begin{eqnarray}
\label{eq:transient}
z(t)=z_0+A\exp(-\gamma t/2)\left[\cos(\omega_0 t)+\frac{\gamma}{2\omega_0}\sin(\omega_0 t)\right],
\end{eqnarray}
where $z_0$ is the equilibrium position and $A$ is the initial displacement from $z_0$.
Figure~\ref{fig:parametermeasure}(d) shows the fitting result with Eq.~(\ref{eq:transient}). For the forced harmonic oscillation method as shown in Eq.~(\ref{eq:analytic2}), the vibrational amplitude displays the Lorentzian shape because $\omega(R)\rightarrow\omega_0$ and $\gamma(R)\rightarrow\gamma$ in the weak driving amplitude $\epsilon$.  In the case of the above methods in Figs.~\ref{fig:parametermeasure}(a) and (b), only the single atomic cloud  oscillates in time. In these situations, because the center of the moving cloud  can be detected reliably through the Gaussian function fitting, the small vibrational amplitude of the oscillators can be resolved well even if the weak modulation amplitude is applied. On the other hand, in the case of parametric resonance, the amplitude of parametrically driven motion becomes the Gaussian-like-shaped function centered at twice the natural frequency $\omega_0$ for the small modulation amplitude $\epsilon$ (Eq.~(\ref{parameter_Duffing})), for which it is almost difficult to separate clearly the two atomic clouds  and rather displays the single merged cloud. Therefore, it is difficult to determine the threshold amplitude $\epsilon_T$ (Eq.~(\ref{eq:characteristicfreq1})) to estimate the damping coefficient while the trap frequency is measured precisely. Nevertheless, if one only wants to estimate roughly the trap frequency, the parametric resonance and the forced harmonic oscillation method are the convenient ways to approximate it by simple observation of the maximum amplitude.

\subsection{Parametrically driven Duffing oscillator (PDDO) and resonantly driven Duffing oscillator (RDDO)}
\label{sec2_nonlineardynamics}
 For the realization of the driven nonlinear oscillators in the MOT, one should modulate periodically the intensities $s_0^{\pm}$ of the two counterpropagating lasers with the relative arbitrary phase given by, 
 \begin{equation}\label{eq:intensity_modulation}
  s_0^{\pm}=s_0 \left[1+\epsilon\cos(\omega_Ft+\phi_{\pm})\right],
 \end{equation}
where $\epsilon$, $\omega_F$ and $\phi_{\pm}$  are the amplitude, the frequency and the phases of intensity modulation, respectively. When the relative phase difference $\Delta\phi=\phi_{+}-\phi_{-}$ becomes zero, the PDDO can be realized \cite{K Kim PRL 2006, K. Kim 2004}. On the other hand, when $\Delta\phi=\pi$, the RDDO can be achieved \cite{gmoon njp2013,gmoon jkps2011}. Then, from  Eq.~(\ref{MOTeq}), one obtains the simple one-dimensional equation of motion in the $z$ direction as,
\begin{subequations}
\begin{eqnarray}
\ddot{z}+\gamma\dot{z}+\omega_{0}^{2}\left(1+\epsilon\cos\omega_F t\right)z+B_{0}\left(z+\kappa\dot{z}\right)^{3}=F_{\mathrm{ext}},~&\text{for the PDDO}&\\
\ddot{z}+\gamma\dot{z}+\omega_{0}^{2}z+B_{0}\left(z+\kappa\dot{z}\right)^3=F_{0}\cos\omega_F t+F_{\mathrm{ext}},~ &\text{for the RDDO}&
\end{eqnarray}\label{nonlineareq}
\end{subequations}
where
\begin{eqnarray}
\label{parameter_Duffing} 
\kappa=\hbar k/(\mu_{B}b),~\quad F_{0}=\frac{\hbar k\Gamma s_{0}\epsilon}{m_a(4(\Delta/\Gamma)^2+1)}, ~~F_{\mathrm{ext}}=F_{\mathrm{sh}}+f(t).
\end{eqnarray}

 The additional term $F_{\mathrm{ext}}$ represents the combined contributions of the atom-atom interaction, so called the shadow force $F_{\mathrm{sh}}$, and the random force $f(t)$ due to the spontaneous emission of a single photon, which satisfies $\left<f(t)f(t')\right>=2\gamma k_B T /m_a \delta(t-t')$ where $\delta(t)$ is the Dirac delta function. These terms cause the cooperative and the stochastic dynamics as will be shown in Sec.~\ref{sec3}$\sim$\ref{sec5}; but in the present section, we ignore these terms and focus on the nonlinear dynamics.
 
\begin{figure}[ht] \centering
\includegraphics[scale=0.35]{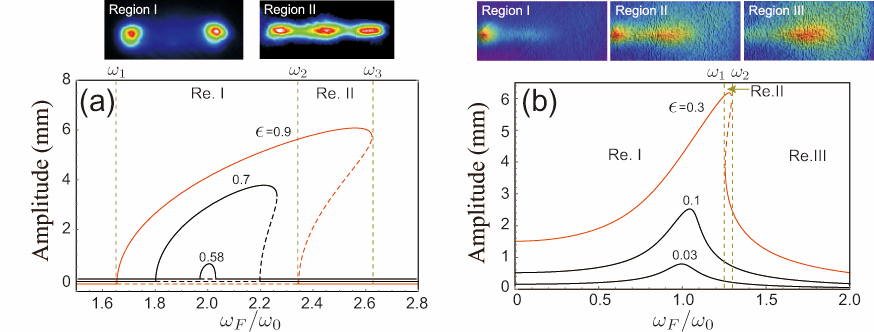}
\caption[figure3]{\label{fig:vibrational motion}\small The vibrational amplitude versus the driving frequency normalized to the trap frequency $\omega_F/\omega_0$ of (a) the parametrically driven Duffing oscillator (PDDO) and (b) the resonantly driven Duffing oscillator (RDDO) depending on the modulation amplitude $\epsilon$. The solid line and dashed line in (a) and (b) stand for the stable state and the unstable state, respectively. The frequency $\omega_1$ and $\omega_2$ of the red curve in (a)  denote the supercritical and subcritical Hopf-bifurcation point, respectively, and each curve is shown with a small offset to distinguish each other. In each region I and II, the atoms are occupied in the stable states as shown in the uppermost picture. In the case of the RDDO (red line) where $\epsilon=0.3$, the uppermost pictures in (b) show the atoms occupied in the stable states depending on the driving frequency $\omega_F$.}
\end{figure}
 
 We now investigate the behavior of each oscillator with respect to the modulation amplitude and frequency. We can assume the steady-state solution of Eqs.~(\ref{nonlineareq})  vibrating at the frequency $\omega_{d}$ as in the following,
 \begin{equation}
 z(t)=R(t)\cos[\omega_d t+\varphi(t)],\label{eq:ansatz_R_phi}
 \end{equation}
where $R(t)$ and $\varphi(t)$ are the slowly varying functions of time.
In the case of the PDDO (Eq.~(\ref{nonlineareq}a)), when the modulation frequency is about twice the trap frequency ($\omega_F\approx 2\omega_0$), the parametric excitation occurs, leading to the development of the period-two state, a state vibrating at twice the modulation period $2\tau_F(\tau_F=2\pi/\omega_F)$. Because the equation of motion Eq.~(\ref{nonlineareq}a) is invariant under a discrete time translation by $t\rightarrow t+\tau_F$, there can exist two period-two states, $z^{(1)}(t)$ and $z^{(2)}(t)$, vibrating out-of-phase satisfying $z^{(1)}(t)=z^{(2)}(t+\tau_F)$~\cite{L. D. Landau 1976}. These two identical states will be discussed more in detail in Sec.~\ref{sec2} in terms of the spontaneous breaking of the discrete time-translation symmetry.  When one inserts Eq.~(\ref{eq:ansatz_R_phi}) with $\omega_d=\omega_F/2$ into Eq.~(\ref{nonlineareq}a) and applies the method of averaging \cite{S. H. Strogatz 2001},  one obtains the two coupled equations, 
\begin{eqnarray}
\label{eq:Solutions1}
\frac{d R}{d t}&=&-\frac{\gamma}{2}R\left[1+\frac{3B_0}{16\omega_F^2}(4+\epsilon_T^2)R^2-\frac{\epsilon}{\epsilon_T}\sin2\varphi\right], ~~ \epsilon_T= 2\kappa\omega_0\quad\nonumber\\
\frac{d \varphi}{d t}&=&-\frac{\omega_F-2\omega_0}{2}+\frac{\epsilon\omega_0}{4}\cos2\varphi+\frac{3B_0}{32\omega_0}(4+\epsilon_T^2)R^2,
\end{eqnarray}
where $\epsilon_T$ is the threshold modulation depth that will be described after Eq.~(\ref{eq:characteristicfreq1}). The steady-state solutions of $R$ can assume a finite value as well as zero. The stabilities of these solutions can change depending on the parameters, leading to the Hopf bifurcation where the stable fixed point makes a transition to the limit cycle~\cite{KihwanPRA:Hopf}. As seen by the red curve in Fig.~\ref{fig:vibrational motion}(a), when $\omega_F$ increases, this transition occurs continuously across $\omega_1$ (supercritical bifurcation) and the limit cycle persists up to $\omega_3$. When we decrease $\omega_F$, there happens a sudden jump to the large-amplitude limit cycle (subcritical bifurcation) at $\omega_F=\omega_2$. We can easily obtain the following relations,  
\begin{eqnarray}
\label{eq:characteristicfreq1}
\omega_{1(2)}&=&2\omega_0-(+)\frac{\omega_0}{2}\sqrt{\epsilon^2-\epsilon_T^2},\nonumber\\
\omega_3&=&\omega_0\left[1+\frac{\epsilon}{2\epsilon_T}\sqrt{4+\epsilon_T^2}\right].
\end{eqnarray}
It can be seen that $\epsilon_T$ is the threshold value of the modulation amplitude $\epsilon$, above which the parametric resonance with the nonzero steady-sate $R$ can occur.  It is noted that there can exist many subharmonic resonances around $\omega_F=2\omega_0 / n$ ($n$: positive integer)~\cite{Noh2006}, which has also been an interesting subject for synchronization of coupled systems~\cite{Razvi1998,Shim2007,Turner1998} 

In the case of the RDDO, Eq.~(\ref{nonlineareq}b) shows the resonant behavior at the condition $\omega_F\approx\omega_0$ and has a solution vibrating at the frequency $\omega_d=\omega_F$. Then the coupled equations for the amplitude $R$ and the phase $\varphi$ of Eq.~(\ref{eq:ansatz_R_phi}) become,
\begin{eqnarray}
\label{eq:Solutions2}
\frac{d R}{d t}&=&-\frac{F_0\sin\varphi}{2\omega_F}-\frac{\gamma}{2} R-\frac{3B_0 C}{8}R^3 , \nonumber\\
R\frac{d \varphi}{d t}&=&-\frac{F_0\cos\varphi}{2\omega_F}+\frac{\omega_0^2-\omega_F^2}{2\omega_F}R+\frac{3B_0 C}{8\kappa\omega_F}R^3,
\end{eqnarray}
where $C\equiv\kappa(1+\kappa^2\omega_F^2)$. The steady-state $R$ satisfies,
\begin{eqnarray}
\label{eq:analytic2}
R&=&\frac{F_0}{\left[(\omega_{F}^2-\omega^2(R))^2+\gamma^2(R)\omega_{F}^2\right]^{1/2}},
\end{eqnarray}
where $\omega(R)=\left[\omega^2_0+3B_0 C R^2/(4\kappa)\right]^{1/2}$ is the eigenfrequency of the nonlinear oscillation and $\gamma(R)=\gamma+3B_0C R^2 /{4}$ is the damping coefficient. 

Figure~\ref{fig:vibrational motion} shows the vibration amplitude, Eq.~(\ref{eq:ansatz_R_phi}), while varying the driving frequency $\omega_F$ for both cases of nonlinear oscillators.  When the driving amplitude $\epsilon$ is above a certain threshold value, there is a  range ($\omega_2<\omega_F<\omega_3$ for the PDDO and $\omega_1<\omega_F<\omega_2$ for the RDDO) of the driving frequency, where three different steady-state---two stable (solid curves) and one unstable (dashed curve)---solutions exist. This indicates that  the system that ends up with within this region is determined by the sweeping direction of $\omega_F$, which may allow one  to observe the hysteresis. For instance, in the case of the PDDO with $\epsilon=0.9$ (Fig.~\ref{fig:vibrational motion}(a)), as $\omega_F$ is increased, the system follows the large-amplitude state until $\omega_F=\omega_2$ and makes a sudden jump to the zero-amplitude state. For the decrease of $\omega_F$, on the other hand, the system stays at the zero-amplitude state until $\omega_F=\omega_1$ and then jumps to the finite-amplitude state. Notice that for our system that consists of cold atoms at a finite temperature, both stable states can be populated due to thermal fluctuations as seen in the top images in the region II of Fig.~\ref{fig:vibrational motion}. 

The two stable states of the RDDO (solid curves in the region II in Fig.~\ref{fig:vibrational motion}(b)) oscillate periodically in time, and thus have the time-translational symmetry $z^{(i)}(t)=z^{(i)}(t+\tau_F)$. The superscript $i$ of $z$ indicates each of stable states $i=1,2$. But there is no symmetry between the two stable states in the RDDO whereas those in the PDDO possess the time-translational symmetry such that $z^{(2)}(t)=z^{(1)}(t+\tau_F)$ as described before. This can be observed experimentally in the uppermost figures of Fig.~\ref{fig:vibrational motion}. Atoms in the PDDO occupy two identical out-of-phase stable states vibrating symmetrically around the trap center. However, the two stable states occupied by atoms in the RDDO have  different vibrational amplitudes. The difference in the symmetry between the PDDO and the RDDO causes the differing responses of the noise-induced switching dynamics with respect to the inter-particle interaction, which will be addressed in Sec.~\ref{sec 3.2} in more detail.
The typical parameters for the PDDO and RDDO in our system are shown in Table.\ref{parameters}.
 
\begin{table}[ht]
\caption{The typical parameters for two nonlinear oscillators}
\label{parameters}
\centering % used for centering table
\begin{tabular}{c c c} % centered columns (4 columns)
\hline\hline %inserts double horizontal lines
Parameters & PDDO & RDDO \\ [0.5ex] % inserts table
%heading
\hline % inserts single horizontal line
$\omega_0$ & $2\pi\times50.1$Hz & $2\pi\times32.7$Hz \\
$\gamma$ & 90.9 s$^{-1}$ & 40.6 s$^{-1}$ \\
$B_0$ & $1.9\times10^9$m$^{-2}$$\cdot$s$^{-2}$ & $9.7\times10^8$m$^{-2}$$\cdot$s$^{-2}$ \\
$\epsilon$ & 0.9 & 0.3 \\
$F_0$ & -- & 66.13 m$\cdot$s$^{-2}$\\ [1ex] % [1ex] adds vertical space
\hline %inserts single line
\end{tabular}
\label{table:nonlin} % is used to refer this table in the text
\end{table}
 
\subsection{Complex nonlinear dynamic behavior in the modulated cold atom system}
\label{magnetic modulation}

 The complex structures of the basin of attraction as well as the strange attractors appear naturally in the nonlinear dynamic systems. Obtaining and analyzing these structures is the basic tool for understanding the behaviors of complex dynamic systems~\cite{Aguirre2009}. In the parametrically modulated MOT system, there exist rich nonlinear dynamic properties such as the periodic doubling and the chaotic motion when one strongly drives the system while the system parameters undergo through the specific values. 
\begin{figure}[ht] \centering
\includegraphics[scale=0.7]{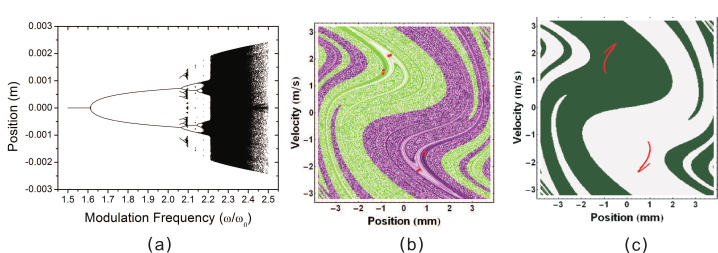}
\caption[figure4]{\label{fig:magnetic modulation}\small (a) Bifurcation map of the parametrically modulated system via the magnetic field-gradient modulation when the modulation amplitude $\epsilon=0.8$. The simulation parameters are $s_0$=0.03, $\Delta$=-2.7$\Gamma$ and $b$=0.15 T/m. Each data point is taken when $t$=400$\times$ 2$\pi$/$\omega_F$. (b) The attractors and their basin of attraction are shown at $t$=400$\times$ 2$\pi$/$\omega_F$ when the modulation amplitude $\epsilon=0.8$ and modulation frequencies $\omega_F$ is 2.15$\omega_0$, which represents the period-8 state. The red dots in the basin of attraction represent the attractors of the corresponding conditions. (c) The Strange attractors and their basins of attraction are shown when  $\omega_F =$ 2.2$\omega_0$. The other parameters are the same as in (b).}
\end{figure}

The modulation of the magnetic field gradient $b$ or the laser detuning $\Delta$ leads to a stronger nonlinear oscillation than that due to the laser intensity modulation because these parameters appear in the denominator of the equation of motion Eq.~(\ref{MOTeq}). Therefore, one can investigate more diverse complex dynamic behaviors beyond the parametric resonance. In the magnetic field-gradient modulation case, when the modulation amplitude is small, the dynamics of atoms exhibits the parametric resonance same as the case of intensity modulation in Sec.\ref{sec2_nonlineardynamics}. However, as the modulation amplitude is larger, one  obtains the complex bifurcation maps and  can find the complicated dynamics such as the chaotic motion and the corresponding basins of attraction in the specific conditions as shown in Fig.~\ref{fig:magnetic modulation}~\cite{Yonghee2011}.

  The most remarkable characteristics of the magnetically modulated atomic system is that the coefficient of the cubic nonlinear term (and also the higher order terms) is modulated at more than two different frequencies. This magnetic modulation provides the bigger nonlinear effects and therefore allows the more complex and richer dynamics than those in the systems for the laser intensity modulation. In experiment, we apply the periodic modulation to the additional Helmholtz coil, whose modulation frequency is tuned carefully to suppress the parametric excitation of the $x$ and $y$ motions. We expect that this complex structure and the intrinsic thermal fluctuations can provide a new possibility to explore the stochastic phenomena occurring between the complex basins of attraction~\cite{Silchenko2003,Beri2008}. Nonetheless it may be hard to observe directly because of some technical challenges such as the imaging resolution and the perturbation due to the eddy current on the vacuum chamber. 
\begin{figure}[ht] \centering
\includegraphics[scale=0.7]{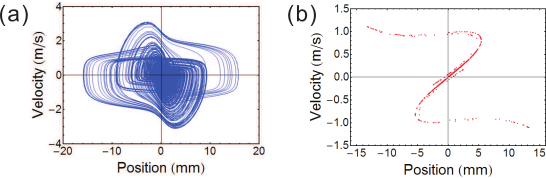}
\caption[figure5]{\label{fig:DM}\small Typical calculation results for modulation of the laser detuning. (a) Trajectories in the phase space and (b) strange attractors in the Poincar\'{e} section.The simulation parameters are $s_0=0.01$, $b=0.10$, $\Delta=-2.4\Gamma$, $\omega_F=2\omega_0$ and $\epsilon=0.9$.}
\end{figure}
 
When one modulates the trap-laser frequency detuning, the system shows the stronger nonlinearity even at a small amount of modulation. For example, the modulation amplitude $\epsilon=0.1$ for the detuning modulation produces almost the same effects as those with $\epsilon=0.95$ for the intensity modulation case. If we modulate the detuning with the amplitude $\epsilon=0.9$, the trap frequency is modulated almost as the delta-kicked rotor, and the damping as well as the nonlinear terms are also strongly modulated simultaneously. The system dynamics then becomes chaotic as shown in Fig.~\ref{fig:DM}, but  one may not observe directly  similarly to the case of the magnetic field gradient modulation.

 In this section, we have briefly introduced the wide variety of the nonlinear dynamic behaviors available in the modulated cold atom system. The complex structure of the nonlinear dynamics is an interesting topic, but in the present review, we focus mainly on the PDDO and the RDDO that have the dynamical bistable states, which allows investigation of the fluctuational dynamics and the cooperative phenomena originating from the atom-atom interaction.

\section{Theoretical description of the noise-induced switching dynamics}
\label{sec3}
In the previous section we have discussed the noise-free single-particle nonlinear dynamics and found the range of the parameters (i.e. the driving frequency and the amplitude) where  the dynamical stable states exist. Specifically, we are interested in the region I in Fig.~\ref{fig:vibrational motion}(a) for the PDDO and the region II in Fig.~\ref{fig:vibrational motion}(b) for the RDDO. Now we will look into the dynamics of atoms between the stable states under the influence of thermal fluctuations. Without the fluctuations, an atom will decay into one of the stable states depending on its initial state, and reside there indefinitely. However, the fluctuations can induce switching between the stable states and the resulting distribution of atoms will be determined by the switching rates, not by the initial conditions any more. A similar example in thermal equilibrium is the damped double well potential where the atomic distribution is governed by the switching rate from each well~\cite{J. Hales 2000,N. G. van Kampen 2007,M. Borkovec 1990,L. I. McCann 1999}, where the switching rate is given by Kramers' rule in the form of, 
\begin{eqnarray}\label{MP14}
W=\mathrm{const}\times\exp(-\mathcal{R}/D). \nonumber
\end{eqnarray}
Here $\mathcal{R}$ is the activation energy or simply the potential barrier and $D$ is the noise intensity. Although, in out-of-equilibrium systems, one cannot  define the activation energy, it has been known that one is still able to find the switching rate in the similar activation form  \cite{M. I. Dykman 5902,M. I. Dykman 747}. It has been also experimentally verified in various systems such as the single electron in a Penning Trap \cite{L J Lapidus PRL 1999}, the micreomechanical systems (MEMS) \cite{Chan PRL 2008,Chan PRL 2007} and the Josephson junctions~\cite{N. Gronbech 2004, Siddiqi  2005}. Sec.~3.1 will provide the theoretical tool to calculate the switching rates of an atom for both types of oscillators. In Sec.~3.2, we will include the atom-atom interactions and show how the switching rates are modified. Finally the master equation governing the time evolution of the system will be derived. Its steady-state solutions determine the distributions of atoms depending on the control parameters, which is the key to understand the collective phenomena as described in Sec.~4 and Sec.~5.

\subsection{Noise-induced switching dynamics}
Here we introduce the method to calculate the activation-form switching rate in a driven system. First we rewrite the equations of motion in the rotating frame. Then, to obtain the activation energy-like quantity, we look for the most probable path of the switching atom from one of the stable states, during which the activation energy can be obtained. 

\subsubsection{Equation of motion in the rotating frame}
\label{the rotating frame}
\begin{figure}[hp] \centering
\includegraphics[scale=0.45]{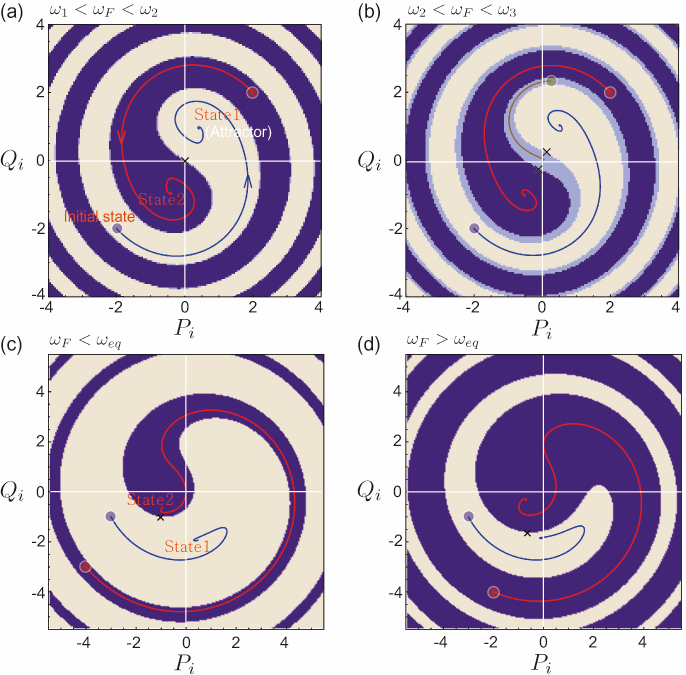}
\caption[figure6]{\label{fig:Typicaltrajectory}\small Typical trajectories, attractors, and their basins of attraction in the rotating frame in the absence of the noise and the interaction,
depending on the driving frequency $\omega_F$. (a) and (b) represent the case of PDDO, while (c) and (d) the case of RDDO. Small colored circles in each figure indicate the initial states of atoms and the colored solid lines the atomic trajectories towards the attractor (stable state) starting from the initial states. The cross in the figures represents the unstable state. $\omega_{eq}$ in (c) and (d) stands for the driving frequency at which the population of each state is equal with each other.}
\end{figure}
Both nonlinear dynamics discussed in Eq.~(\ref{nonlineareq}) can be better described in the rotating frame, which provides one the approachable static picture, using the standard transformation, as given by~\cite{M. I. Dykman 5902},
\begin{eqnarray}
z_i&=&C_{\mathrm{RWA}}\left[P_i\cos(\omega_d t)-Q_i\sin(\omega_d t)\right],\nonumber\\
\dot{z}_i&=&-\omega_d C_{\mathrm{RWA}}\left[P_i\sin(\omega_d t)+Q_i\cos(\omega_d t)\right],
\end{eqnarray}
where the subscript $i$ denotes the $i$th atoms. The equations of motions for the slow variables $\mathbf{q}_i\equiv(Q_i,P_i)$ with respect to the slowly varying time $\tau$ are then given by,
\begin{eqnarray}\label{rotating}
&&\frac{d\mathbf{q}_i}{d\tau}=\mathbf{K}_i(\mathbf{q}_i)+\mathbf{f'}(\tau),~\mathbf{K}_i(\mathbf{q}_i)=-\zeta^{-1}\mathbf{q}_i+\hat{\epsilon}\partial_{\mathbf{q}_i}H_\mathrm{tot}=\mathbf{K}^{(0)}_i(\mathbf{q}_i)+\hat{\epsilon}\partial_{\mathbf{q}_i}H_{\mathrm{sh}},
%&&\mathbf{K}^{(0)}_i(\mathbf{q}_i)=-\zeta^{-1}\mathbf{q}_i+\hat{\epsilon}\partial_{\mathbf{q}_i}H^{(0)}(\mathbf{q}_i),
\end{eqnarray}
where $H_\mathrm{tot}=H^{(0)}(\mathbf{q}_i)+H_\mathrm{sh}$ and the tensor $\hat{\epsilon}$ is the permutation tensor; $\epsilon_{Q_iQ_i}=\epsilon_{P_iP_i}=0$ and $\epsilon_{Q_iP_i}=-\epsilon_{P_iQ_i}=1$.
The generalized force $\mathbf{K}_i(\mathbf{q}_i)$ acting on the $\emph{i}$-th atom depends on the dynamical variables $\mathbf{q}_i$ of all the particles, and
the terms $\mathbf{K}^{(0)}_i(\mathbf{q}_i)$ and $H_{\mathrm{sh}}$ in Eq.~(\ref{rotating}) represent the force in the absence of the interaction and the atom-atom attractive potential, respectively.  In Sec.~\ref{sec 3.2}, we discuss the switching rate modification due to the attractive interaction potential $H_{\mathrm{sh}}$ in detail, but in this section we neglect $H_{\mathrm{sh}}$ for now. In this case, the time evolution of the system is dominated by the effective hamiltonian $H^{(0)}(\mathbf{q}_i)$  given by,
\[ H^{(0)}(\mathbf{q}_i)=
  \begin{cases}
    \frac{1}{4}\left(Q_i^2+P_i^2\right)^2+\frac{1}{2}\left(1-\mu\right)P_i^2-\frac{1}{2}\left(1+\mu\right)Q_i^2~,  & \text{PDDO}\\
    \frac{1}{16}\left(Q_i^2+P_i^2\right)^2-\frac{1}{2}\left(Q_i^2+P_i^2\right)-\frac{F_0}{2C_{\mathrm{RWA}}\omega_F|\delta\omega|}P_i~,  & \text{RDDO}\\
  \end{cases}
\]
where $i=1,\ldots,N_{\mathrm{tot}}$, $\partial_{\mathbf{q}_i}\equiv(\partial_{Q_i},\partial_{P_i})$ and $\delta\omega=(\omega_F-\omega_0)$. Here $\mathbf{f'}(\tau)$ is the white Gaussian noise with the two asymptotically independent components,
\begin{eqnarray}
&&\left<f'_j(\tau)f'_{j'}(\tau')\right>=2D_{\tau}k_BT\delta_{jj'} \delta(\tau-\tau')\label{fluctuation1},
\end{eqnarray}
where $k_B$ and $T$ are the Boltzmann constant and the temperature, respectively.  The various parameters such as $C_{\mathrm{RWA}}$, $\omega_d$, $\mu$, $D_{\tau}$, $\zeta$ and $\tau$ are listed in Table~\ref{parameters1}.
\begin{table}[ht]
\caption{The various parameters such as $C_{\mathrm{RWA}}$, $\omega_d$, $\mu$, $D_{\tau}$, $\zeta$, and $\tau$}
\label{parameters1}
\centering % used for centering table
\begin{tabular}{c c c c c c c} % centered columns (4 columns)
\hline\hline %inserts double horizontal lines
Oscillators & $C_{\mathrm{RWA}}$ & $\omega_d$ &$\mu$&$D_{\tau}$&$\zeta$&$\tau$\\ [0.5ex] % inserts table%heading
\hline % inserts single horizontal line
PDDO & $(2\epsilon\omega_0^2/3|B_0|)^{1/2}$ & $\frac{\omega_F}{2}$&$\frac{\omega_F(\omega_F-2\omega_0)}{2\epsilon\omega_0^2}$&$\frac{6B_0\gamma}{\omega_0^4\omega_F\epsilon^2m_a}$&$\frac{\epsilon\omega_0^2}{\omega_F\gamma}$&$\frac{\epsilon\omega_0^2}{2\omega_F}t$ \\
RDDO &$(2\omega_F|\delta\omega|/3|B_0|)^{1/2}$&$\omega_F$&$0$&$\frac{3B_0}{2\omega_F^3\gamma m_a}$&$\frac{2|\delta\omega|}{\gamma}$&$|\delta\omega|t$\\[1ex]% [1ex] adds vertical space
\hline %inserts single line
\end{tabular}
\label{table:nonlin} % is used to refer this table in the text
\end{table}
Figure~\ref{fig:Typicaltrajectory} shows the trajectories, attractors and their basins of attraction in the rotating frame in the absence of the noise and the interaction, corresponding to the PDDO ((a) and (b)) and the RDDO ((c) and (d)).

As is well known, the attractor represents a set of states towards which the neighboring states in a given basin of attraction  approach asymptotically in the course of dynamic evolution. The small colored circle and the cross in each figure represent the initial state and the unstable state, respectively, and the colored solid curves are the trajectories of an atom that evolves to the attractor (stable state). In Fig.~\ref{fig:Typicaltrajectory}(c) and (d), the stable state 1(2) indicates the large(small)-amplitude state. Assuming no fluctuations in the atomic motion, the initial conditions of the atomic position and velocity determine the attractors where the atom ends up. For example, if the initial condition of an atom lies in a certain colored region, the atom approaches the attractor located in the same basin. In reality, however, atomic fluctuating motions exist due to spontaneous emission resulting in the broadened distributions of atomic position and velocity near the stable attractors. Due to the large diffusion of atomic motion resulting from spontaneous emission, certain atoms may jump far from the original attractor and be transferred to another attractor through the unstable regions near the boundary. For this case, the shape of the atomic phase-space distribution in each nonlinear oscillator is shown in the uppermost picture of Fig.~\ref{fig:vibrational motion}, depending on the driving frequency.

\subsubsection{Calculation of the switching rate}
We are interested in the atomic dynamics due to the thermal noise ${\bf f'}(\tau)$ with $H_{\mathrm{sh}}=0$ in Eq.~(\ref{rotating}), which adds the fluctuations to the stable states and leads to the transition between the two states. A question now arises: how can we describe this random dynamics systematically? The appropriate theory is based on the random fluctuations and the optimal path. The main idea is that, although the motion of system is random due to fluctuations, which allows the atomic escape from the stable state, the system is most likely to move along the particular trajectory known as the optimal path~\cite{M. I. Dykman 747,M. I. Dykman 587,F. Moss vol1}.

The most suitable theoretical approach to the fluctuation-induced transition (or escape) problem is based on the path-integral formalism~\cite{Feyman 1965}. The optimal path gives rise to a maximum in the prehistoric probability density distribution, $p_h({\bf q},\tau | {\bf q}_f,\tau_f)$~\cite{M. I. Dykman 587,M. I. Dykman 5229}, which is the conditional probability density for a system arriving at the point ${\bf q}_f$ at the time $\tau_f$ to have passed through the point ${\bf q}$ at the time $\tau$. The form of $p_h$ can be then expressed in the path integral relation~\cite{M. I. Dykman 587},
 \begin{eqnarray}
p_h({\bf q},\tau|{\bf q}_f,\tau_f)&=&C \int_{{\bf q}(\tau_i)\approx{\bf q}^{(0)}(\tau_i)}^{{\bf q}(\tau_f)={\bf q}_f} \mathcal{D} {\bf q}(\tau') \delta({\bf q}(\tau)-{\bf q}) \nonumber \\
&&\times \exp\left[-\frac{\mathcal{R}[{\bf q}(\tau)]}{D}-\frac{1}{2}\int_{\tau_i}^{\tau_f} d\tau' \frac{\partial \mathbf{K}(\mathbf{q})}{\partial {\bf q}}\right],~~ \tau_i \rightarrow -\infty,~
\end{eqnarray}
where $C$ is the normalization constant determined by the conditions $\int d{\bf q}~p_h =1$ and $D=D_\tau k_B T$. $\mathcal{R}[{\bf q}(\tau)]$ is the action functional for an auxiliary dynamical system with the Lagrangian $L$,
 \begin{eqnarray}
 \label{lagrangian}
\mathcal{R}[{\bf q}(\tau)]=\int_{t_i}^{t_f} d\tau L(\dot {\bf q}, {\bf q} ;\tau),~L=\frac{1}{4}(\dot {\bf q} -{\bf K})^2.
\end{eqnarray}

\begin{figure}[ht]\centering
\includegraphics[scale=0.5]{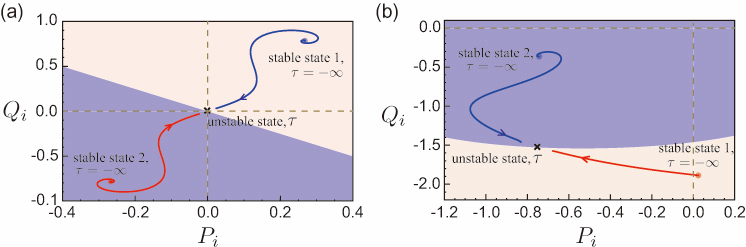}
\caption[figure7]{\label{fig:optimal path}\small The optimal path (or the most probable escape path) of the fluctuation-induced transition for (a) the PDDO and (b) the RDDO, which starts from the attractor position at $\tau=-\infty$ and arrives at the origin at the finite $\tau$. The phase space area of (a) and (b) is an enlarged view around the unstable state in Fig.~\ref{fig:Typicaltrajectory}(a) and in Fig.~\ref{fig:Typicaltrajectory}(d), respectively.}
\end{figure}

In the range of small noise intensity $D$, the optimal path (where $p_h$ has a peak value) is given by the condition that the action $\mathcal{R}[{\bf q}]$ be minimum,
\begin{equation}
\label{variation}
\mathcal{R}_n=\min \int_{t_i}^{t_f} d\tau L(\dot{\bf q}, \bf q ;\tau).
\end{equation}
The variational problem of Eq. (\ref{variation}) for $\mathcal{R}[{\bf q}]$ can be solved with the Hamiltonian equations of motion of an auxiliary system with Eq.~(\ref{lagrangian}),
\begin{eqnarray}
\label{H-Jeqs}
&&H={\bf p}^2+{\bf p}\cdot{\bf K},~~\text{where}~~ {\bf p}=\frac{1}{2} (\dot{\bf q}-{\bf K}),\\
&&\dot{\bf q}={\bf K}+2{\bf p}, ~~\dot{\bf p}=-{\bf p}\cdot \partial_{\bf q} {\bf K},~~\dot{\mathcal{R}}= {\bf p}^2,\nonumber
\end{eqnarray}
with the boundary conditions that the motion starts at $\tau \rightarrow-\infty$ in the initially occupied stable state ${\bf q}_n (\tau)$ and asymptotically approaches the saddle point at $\tau\rightarrow \tau_f$. One can solve the variational equations Eq.~(\ref{variation}) with Eqs.~(\ref{H-Jeqs}) numerically, and then  obtain the $\mathcal{R}_n$, so called the \textit{activation energy}, as well as the optimal path ${\bf q}_{\mathrm{opt}}$~\cite{M. I. Dykman 747}. In fact, $\mathbf{q}_{\mathrm{opt}}$ determines the most probable path that the particle follows during switching, which is displayed in Fig.~\ref{fig:optimal path}.

The probability of the noise-induced switching from the $n$th to the $m$th state is determined by the activation energy $\mathcal{R}_n$ and the noise intensity $D$ as follows,
\begin{eqnarray}
\label{transitionprobability}
&&W_{nm}=C_{W}\exp\left(-\frac{\mathcal{R}_n}{D}\right),
\end{eqnarray}
where the prefactor $C_W \sim \textnormal{max}(\omega_{cl},\gamma / 2)$, and $\omega_{cl}$ is the frequency of the small-amplitude damped atomic vibration with respect to the atomic cloud center.
The investigation on the optimal path and the activation energy represents an intriguing and important problem in nonequilibrium dynamics.  In the microelectromechanical system, for instance, the lack of the time-reversal symmetry has been observed experimentally~\cite{Chan PRL 2008}.

\subsubsection{Measurement of the switching rate }
\label{sec 3.1.3}
\begin{figure}[hp] \centering
 \includegraphics[scale=0.5]{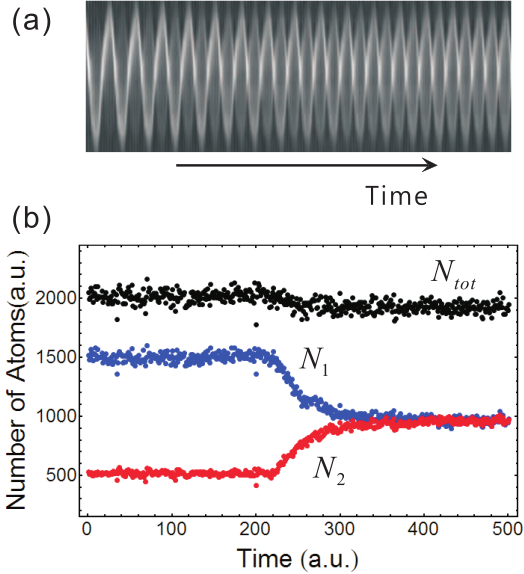}
 \caption[figure8]{\label{fig:transition}\small Typical data of the fluctuation-induced transition. (a) The sequence of the images during the transition. While there is  only one oscillating cloud in the beginning (left), the population of the other cloud is gradually increasing with time until the populations of two clouds become equal. (b) Typical temporal data for measuring the transition rate. $N_1$ and $N_2$ are the populations of each cloud 1 and 2, such that $N_{\mathrm{tot}}=N_{1}+N_{2}$.}
\end{figure}
We have experimentally measured the fluctuation-induced switching rate for atomic transition occurring in the parametrically modulated MOT~\cite{K. Kim PRA 2005}, which was done by emptying one of the two dynamic attractors and by measuring the subsequent repopulation of the empty attractor (or equivalently the depopulation of the populated attractor) with time. This corresponds to the basic experiment for understanding the stochastic and nonequilibrium properties of the system.

To do this, we have utilized two different methods. One is to use the resonant laser light that is cylindrically focused at 5 mm away from the center of the limit-cycle motion. We  turned on the laser light  for 3 ms, and one can selectively blow away one atomic cloud. We then  took the images until the two oscillating clouds have the same populations. The other is to use the additional bias field as described in Sec.~\ref{Response to the symmetry breaking field}. We applied the strong DC bias field in one direction so that all the atoms  occupied only one of the two stable states. This produces the same net effect of blowing away one cloud. We then suddenly turned off the bias field and subsequently measured the transition time. The typical data of the fluctuation-induced transition observed in the parametrically modulated MOT are shown in Fig. \ref{fig:transition}. The transition rate is extracted from the decay of the population difference of the atomic clouds,
\begin{equation}
 N_2 -N_1 = N_0 \exp(-2 W t),
\end{equation}
where $N_1$ and $N_2$ are the population of  each cloud, $N_0$ is the constant and $W$ is the transition rate between two clouds. The measured transition rate $W$ is from 0.1 to 5 s$^{-1}$, depending on the experimental parameters.

\subsection{Noise-induced switching dynamics under the inter-particle interaction}
\label{sec 3.2}
When the atom-atom interactions come into play, one can observe the interesting collective phenomena such as the spontaneous symmetry breaking~(Sec.~\ref{sec4}) and the kinetic phase transition~(Sec.~\ref{sec5}). In this section, we discuss how the atomic interactions modify the single-particle switching rate and consequently produce the collective phenomena.

\subsubsection{Light-induced atom-atom interaction in the rotating frame}\label{subsec:interaction}
\begin{figure}[ht] \centering
 \includegraphics[scale=0.6]{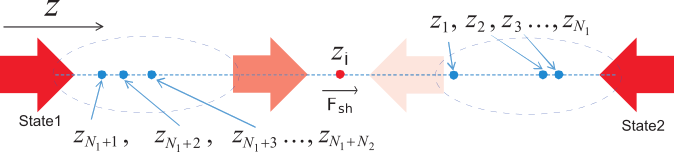}
 \caption[figure9]{ \label{fig:shadow_scheme}\small Modeling of the shadow force in one-dimensional picture. The shielding of atoms from the laser light by the other atoms is described by the sgn-function, which is independent of the distance between the atoms.}
\end{figure}

To explain the collective phenomena in the nonlinear dynamic system consisting of many atoms, one needs to introduce the inter-atomic interactions. As drawn schematically in Fig.~\ref{fig:shadow_scheme}, the interaction on the $i$th atom comes from the shadow force due to the unbalanced absorption of the counterpropagating light. Such an imbalance is associated with \textit{shielding} of the atoms in the stable state since the $i$th switching   atom moves far away from the atomic clouds belonging to each stable state having the atomic number of $N_1$ and $N_2$~\cite{A. M. Steane 1992,Sesko1991,Walker1990}.
In one-dimensional picture, the force on the $i$th atom at the coordinate $z_i$ by the other atoms can be approximately modeled as $F^{i}_{\rm sh}=-f_{\rm sh}\sum_j{\rm sgn}(z_i-z_j)$ as shown in Fig. \ref{fig:shadow_scheme}. This force is weak, much smaller than the Doppler force that confines the atoms in the trap, so that multiple scattering of light can be neglected.

\begin{figure}[hp] \centering
 \includegraphics[scale=0.35]{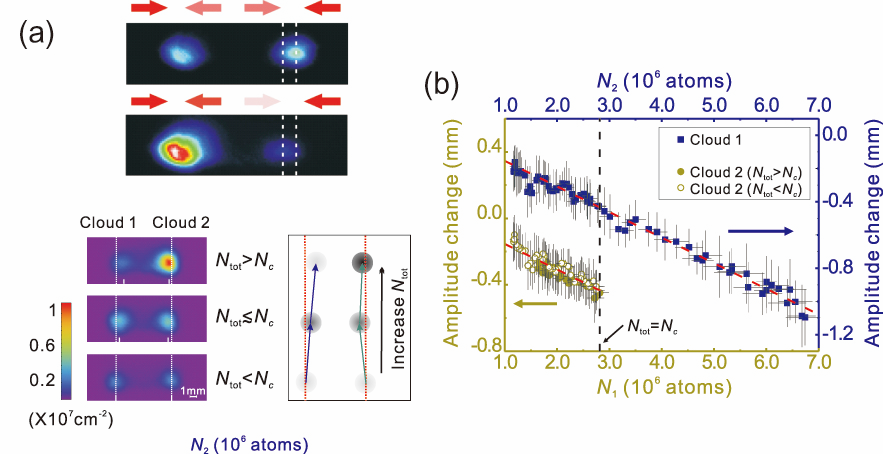}
 \caption[figure10]{\label{fig:shadow}\small (a) Experimental measurement of the shadow force can be made by using the displacements of atomic clouds. The CCD images of the vibrating clouds are taken at the maximal spatial separation ($\approx$ 4.5 mm). In the top panel, the center of the  cloud on the right lies on the white dashed line. It is more attracted toward the larger cloud on the left because the imbalance of the radiation force on the right cloud, due to attenuation of the trapping light, gets bigger as depicted by the red arrows. This can be interpreted as the fact that each of atoms in the right-side cloud is attracted by the mean attractive force contributed by all the atoms in the left-side cloud with respect to those in the right-side one . The mean force, therefore, depends on the number of atoms in the right-side cloud. This behavior can be clearly seen in the bottom panel. The center of the cloud 1 is shifted more to the right side with the increase of the number of atoms in the cloud 2. The typical experimental parameters are $I=0.04 I_s$ and $\Delta=-2.3\Gamma$. The lower, middle, and upper images refer to the symmetric, close to critical, and broken-symmetry states, with $N_{\rm tot}\approx1.5\times10^6$, $5.2\times10^6$, and $6.7\times10^6$ , respectively, while $N_c \approx 5.6\times 10^6$. The cloud centers are indicated by the ticks at the bottom of the CCD images and are schematically drawn in the right-side panel. (b) The displacement of the center of one cloud is measured as a function of the number of atoms in the other cloud. $N_1$ and $N_2$ increase together when more atoms are trapped in the MOT. When the total number $N_{\rm tot}=N_1+N_2$ of atoms reaches a critical value $N_c$ (the vertical dashed line), the spontaneous symmetry breaking occurs as is described in detail in Sec.~\ref{sec4} whereas $N_1$ starts to decrease. Thus the center of cloud 2 is shifted back to the initial value (refer to the bottom panel of (a)). The red dashed lines show the expected linear dependence of the theory.  The error bars show the standard deviations for 20 measurements. (Figures from Ref.~\cite{M Heo PRE 2010})} 
\end{figure}

To estimate the value of $f_{\rm sh}$ that acts on a switching atom, one has to take into account the fact that the atomic clouds are indeed three-dimensional. We consider a simple model in which the laser beam propagating along the $z$-axis passes through the atomic cloud having the density distribution $\rho({\bf r})$ (Fig. \ref{fig:shadow}(a)). The resulting change of the beam intensity $I$ as a function of the transverse coordinates $x,y$ is $\Delta I(x,y)=I\sigma_L\int dz \rho({\bf r})$, where the integration is done over the length of the cloud and $\sigma_L$ is the absorption cross-section. This cross-section depends in the standard way on the intensity $I$ and the frequency detuning. Generally, because of the magnetic-field-induced frequency shift and the Doppler shift, $\sigma_L$ oscillates in time, while the light intensity also oscillates in time. If we disregard these oscillations, for simplicity, we obtain $\sigma_L\approx 5.6\times 10^{-15}$~m$^2$  for typical experimental conditions.
The atomic density distribution $\rho({\bf r})$ can be assumed Gaussian with the same width $w_t\approx 1$~mm in all directions, which was achieved experimentally by tuning the transverse beam intensities. According to the optimal path picture,  atoms most likely move along the MOT axis during switching, and for those atoms the light-intensity change is determined by $\rho({\bf r})$  on the axis. The extra force on the switching atom as it moves between the clouds is then directed toward the more populated cloud and is equal to $f_{\rm sh}|N_1-N_2|$, where, according to the above arguments,
\begin{eqnarray}
\label{shadow_force} f_{\rm sh}=\hbar k \Gamma \sigma_L s_0/4\pi (1+s_0+(2\Delta/\Gamma)^2) w_t^2.
\end{eqnarray}
Here, $k$ is the photon wave number and $s_0=I/I_s$. For $I=0.04 I_s$ and $\Delta=-2.3\Gamma$, we obtain $f_{\rm sh}\approx 2.5\times 10^{-32}$~N. Note that, since the clouds are oscillating, the toatl force $\propto f_{\rm sh}$ is also oscillating, and its time dependence is determined by the sign-function, $F^{i}_{\rm sh}=-f_{\rm sh}\sum_j{\rm sgn}(z_i-z_j)$.

As mentioned in Eq.~(\ref{nonlineareq}), we consider the shadow force on the $i$th atom $F^{i}_{\rm sh}$ in the nonlinear equation. Transferring it to the rotating frame, one can obtain the attractive interaction potential $H_{\rm{sh}}$ in Eq.~(\ref{rotating}).
The Hamiltonian $H_{\mathrm{sh}}$ that describes the shadow-effect-induced interaction then has the form,
\begin{eqnarray}\label{interaction}
H_{\mathrm{sh}}=\frac{1}{2}\sum_{i,j=1}C_{\mathrm{sh}}|{\bf q}_i-{\bf q}_j|,
\end{eqnarray}
where $i=1,~.~.~.~,N_{\mathrm{tot}}$ and
\[ C_\mathrm{sh}=
  \begin{cases}
    \frac{8f_{\mathrm{sh}}}{\pi m_a \omega_0^2 \epsilon C_{\mathrm{RWA}}}~,  & \text{PDDO}\\
    \frac{4f_{\mathrm{sh}}}{\pi m_a \omega_F |\delta\omega| C_{\mathrm{RWA}}}~,  & \text{RDDO}\\
  \end{cases}
\]
Thus the interaction due to the shadow effect is described by the simple Hamiltonian describes in the slow variables.

Though the theoretical description using the shadow-force model is in good agreement with the experimental results  shown in Sec.~\ref{sec4} and~\ref{sec5}, one still needs to check whether the shadow force is the real physical force, not just a model parameter. Measuring the magnitude of the shadow force is also important for calibrating the theoretical description. Figure~\ref{fig:shadow}(a) shows how to measure the shadow force quantitatively in the PDDO which exhibits the time-symmetry breaking. Under the shadow force, the vibrational amplitude of the atomic clouds slightly changed from the initial values. Hence one can measure the amplitude change of the atomic clouds while varying the total number of atoms to analyze quantitatively the shadow force. As shown in Fig.~\ref{fig:shadow}(b), the experimental value of the shadow force  is estimated by the slope of the graph, which yields $f_{\rm sh}\approx 1.86\times 10^{-32}$~N.

\subsubsection{Modification of the switching rate }
\label{Switching rate modification}
The theory on the modification of the switching rate due to the atomic interactions for the PDDO is described in Ref.~\cite{M Heo PRE 2010}. Now we intend to extend this approach to describe the RDDO case. The interaction between atoms leads to an extra force $F^{i}_{\rm sh}=-f_{\rm sh}\sum_j{\rm sgn}(z_i-z_j)$. This force is weak in the sense that it  affects only slightly the inter-cloud atomic dynamics. As reported in Refs.~\cite{M Heo PRE 2010,Dykman PRE 2006}, the modification of the activation energy by the interatomic force on an atom in the cloud $n$ can be written as, to the first order in $H_{\mathrm{sh}}$, 
\begin{eqnarray}
\mathcal{R}_{n} \approx \mathcal{R}_n^{(0)}+\int^\infty_{-\infty}d\tau{\bm \lambda}_n^{\rm (opt)}(\tau)\hat{\epsilon}\partial_{\mathbf{q}}H_{\mathrm{sh}},~
{\bm \lambda}_n^{\rm (opt)}(\tau)=-\frac{1}{2}\left[\dot{\mathbf{q}}_{\mathrm{opt}}-\mathbf{K}^{(0)}_{\mathrm{opt}}\right],
\end{eqnarray}
where $\mathcal{R}_n^{(0)}$ and $\mathbf{q}_{\mathrm{opt}}\equiv\mathbf{q}_{\mathrm{opt}}(\tau)$ is the activation energy without the interaction and the optimal path for the $\emph{i}$th atom in the absence of the interatomic interaction, respectively, and $\mathbf{K}^{(0)}_{\mathrm{opt}}\equiv\mathbf{K}^{(0)}(\mathbf{q}_{\mathrm{opt}}(\tau))$.
Here, $\dot{\mathbf{q}}_{\mathrm{opt}}-\mathbf{K}^{(0)}_{\mathrm{opt}}$ corresponds to the solutions of the variational problem of minimizing the functional $\mathcal{R}[{\bf q}(\tau)]$ in Eq.~(\ref{lagrangian}).

By assuming that all the other atoms stay close to either of the attractors, the total activation energy $\mathcal{R}_{n} (n=1,2)$  can be approximately written as, while taking into account the interaction between atoms,
\begin{eqnarray}\label{activation}
\mathcal{R}_{n}=\mathcal{R}_n^{(0)}+\mathcal{R}^{(1)}_n,~\mathcal{R}^{(1)}_n=\sum_{m=1,2}\alpha_{nm}N_m,
\end{eqnarray}
where $N_n$ is the number of atoms in the cloud $n$ and $\alpha_{nm}$ is given by the explicit expressions,
\begin{eqnarray}
\label{alpa}
&&\alpha_{nn}=\frac{C_\mathrm{sh}}{4}\int^{\infty}_{-\infty}dt\left[\left(\dot{P}^{(n)}_{\mathrm{opt}}-K^{(0)}(P)\right)\aleph_{Q}^{(n)}(t)\right.\nonumber\\
&&~~~~~~~~~~~~~~~~~~~\left.-\left(\dot{Q}^{(n)}_{\mathrm{opt}}-K^{(0)}(Q)\right)\aleph_{P}^{(n)}(t)\right],\nonumber\\
&&\alpha_{n3-n}=\frac{C_\mathrm{sh}}{4}\int^{\infty}_{-\infty}dt\left[\left(\dot{P}^{(n)}_{\mathrm{opt}}-K^{(0)}(P)\right)\tilde{\aleph}_{Q}^{(n)}(t)\right.\nonumber\\
&&~~~~~~~~~~~~~~~~~~~\left.-\left(\dot{Q}^{(n)}_{\mathrm{opt}}-K^{(0)}(Q)\right)\tilde{\aleph}_{P}^{(n)}(t)\right].
\end{eqnarray}
Here
\begin{eqnarray}
\label{alpa1}
&&\aleph_{Q}^{(n)}(t)=\frac{Q^{(n)}_{\mathrm{opt}}(t)-Q^{(n)}_{\mathrm{eq}}}{\sqrt{\xi^{(n)}_1(t)}},\nonumber\\
&&\tilde{\aleph}_{Q}^{(n)}(t)=\frac{Q^{(n)}_{\mathrm{opt}}(t)-Q^{(3-n)}_{\mathrm{eq}}}{\sqrt{\xi^{(n)}_2(t)}},
\end{eqnarray}
where $\aleph_{P}^{(n)}$ and $\tilde{\aleph}_{P}^{(n)}$ are obtained by exchanging $P$ and $Q$  in $\aleph_{Q}^{(n)}$ and $\tilde{\aleph}_{Q}^{(n)}$.
The denominators in Eq.~(\ref{alpa1}) are given by,
\begin{eqnarray}
\label{alpa2}
&&\xi^{(n)}_1(t)=\left(P^{(n)}_{\mathrm{opt}}(t)-P^{(n)}_{\mathrm{eq}}\right)^2+\left(Q^{(n)}_{\mathrm{opt}}(t)-Q^{(n)}_{\mathrm{eq}}\right)^2,\nonumber\\ &&\xi^{(n)}_2(t)=\left(P^{(n)}_{\mathrm{opt}}(t)-P^{(3-n)}_{\mathrm{eq}}\right)^2+\left(Q^{(n)}_{\mathrm{opt}}(t)-Q^{(3-n)}_{\mathrm{eq}}\right)^2.\nonumber
\end{eqnarray}
$\mathcal{R}^{(1)}_n$ in Eq.~(\ref{activation}) shows that the effective  activation energy for switching  depends linearly on the number of atoms in the clouds and $\alpha_{nm}$ for a weak interatomic coupling.
\begin{figure}[thb] \centering
\includegraphics[scale=0.45]{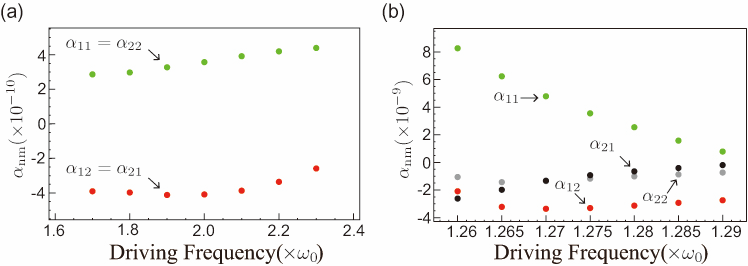}
\caption[figure11]{\label{Alpha}\small Calculation of $\alpha_{nm}$  for (a) the PDDO and (b) the RDDO. Equation~(\ref{alpa}) indicates that the optimal path and the equilibrium position of an atom is critical to determine the value of $\alpha_{nm}$. As a result, $\alpha^{12}=\alpha^{21}$ and $\beta^{12}=\beta^{21}$ for (a), whereas $\alpha^{12}\neq\alpha^{21}$ and $\beta^{12}\neq\beta^{21}$ for (b).}
\end{figure}

Note that $\alpha_{nm}$ depends on the optimal path and the equilibrium position, which is determined by the specific type of nonlinear oscillators under consideration. For instance, the two nonlinear dynamic systems, PDDO and RDDO, have two bistable states but its characteristics are very different. As shown in Fig.~\ref{fig:Typicaltrajectory} and Fig.~\ref{fig:optimal path}, the two dynamical bistable states are symmetric around the rotating-frame phase-space center $(P_i,Q_i)=(0,0)$ in the case of PDDO, whereas that of the RDDO is asymmetric. In particular, the unstable state of the PDDO always is centered at $(P_i,Q_i)=(0,0)$, but that of the RDDO is not at the center position. The differing  characteristics  can give rise to numerous quantitative and qualitative distinctions on the effect of interatomic interaction between the two bistable states. This is because the optimal path of the PDDO is symmetric whereas that of the RDDO is not, so that $\alpha_{nm}$ becomes $\alpha_{12}=\alpha_{21}, \alpha_{11}=\alpha_{22}$ for the PDDO, but $\alpha_{12}\neq \alpha_{21}, \alpha_{11}\neq \alpha_{22}$ for the RDDO, which depends on the modulation frequency $\omega_F$, as shown in Fig.~{\ref{Alpha}}.

According to Eq.~(\ref{activation}) one can show that the activation energy of each stable state changes with the increase of the total number of atoms (i.e., the enhancement of the magnitude of the atom-atom interaction). In other words, even though $|R^{(1)}_n|$ is small compared to the single-atom activation energy $R^{(0)}_n$, the product $\alpha_{nm}N_m (N_{1,2}\gg1)$ can greatly exceed $D$ because the net effect on the switching particle by all the particles in the attractors  can be appreciable. Through Eqs.~(\ref{transitionprobability}), (\ref{activation}), and (\ref{alpa}) one can obtain the switching probability, including the inter-particle interaction, as follows,
\begin{eqnarray}
&&W_{nm}(N_n;N_{\mathrm{tot}})\nonumber\\
&&~~~=W^{(0)}_{nm}\exp[(\alpha^{nm}+\beta^{nm})N_{\mathrm{tot}}-2\alpha^{nm} N_n],
\end{eqnarray}
and
\begin{eqnarray}\label{intertransition}
&&W^{(0)}_{nm}=C_W\exp(-R^{(0)}_{n}/D),\nonumber\\
&&\alpha^{nm}=(\alpha_{nn}-\alpha_{nm})/2D,\nonumber\\
&&\beta^{nm}=-(\alpha_{nn}+\alpha_{nm})/2D.
\end{eqnarray}
As shown in Eq.~(\ref{intertransition}), the distinction of $\alpha_{nm}$ for the two nonlinear oscillators leads to the unique characteristics of the noise-induced switching associated with the inter-particle interaction. In the following section one can observe more obvious differences between them when one considers the master equation.
\subsubsection{The master equation}
\label{masterequation}
To describe the  noise-induced switching between the bistable states, consisting of many particles, one needs to solve the master equation describing the discrete jump processes. Using the steady-state solutions of the master equation, one can calculate the occupied population of each state as in the followings.

\begin{figure}[ht] \centering
\includegraphics[scale=0.6]{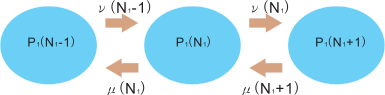}
\caption[figure12]{\label{fig:master}\small Schematic of the discrete jump processes describing the master equation for the time-dependent probability $P_1(N_1,t)\equiv P_1(N_1)$ of state 1, which is the probability of having $N_1$ atoms in the cloud (stable state) 1 at time $t$.}
\end{figure}
Figure~\ref{fig:master} depicts a schematic of the discrete jump processes that leads to derivation of the master equation~\cite{Gardiner} for the time-dependent probability $P_1(N_1,t)$ of the state 1,
where $N_n$ and $N_{\mathrm{tot}}$ is the number of particles in  the state $n(=1,2)$ and the total number of atoms, respectively. The master equation for the time-dependent probability $P_1(N_1,t)\equiv P_1(N_1)$ of the state 1, which corresponds to the probability of having $N_1$ atoms in the cloud (state) 1 at time $t$, is then given by,
\begin{eqnarray}
\label{derivalprodistribution}
&&\partial_tP_1(N_1)=-[\mu(N_1)+\nu(N_1)]P_1(N_1)\nonumber\\
&&~~~~~~~~~~~~~~~+\nu(N_1-1)P_1(N_1-1)+\mu(N_1+1)P_1(N_1+1),~~
\end{eqnarray}
where $\mu(N_1)=N_1W_{12}(N_1;N_{\mathrm{tot}})$ and  $\nu(N_1)=(N_{\mathrm{tot}}-N_{1})W_{21}(N_{\mathrm{tot}}-N_{1};N_{\mathrm{tot}})$.
Here $W_{nm}(N_n;N_{\mathrm{tot}})$ is the transition probability per unit time for one atom in the state $n$ associated with the $n\rightarrow m$ transition, and thus for $N_n$ atoms in the state $n$, the total transition probability per unit time becomes $N_nW_{nm}(N_n;N_{\mathrm{tot}})$. Therefore, in the case where atoms transit from state 1 to state 2, the total transition probability is expressed by $\mu(N_1)=N_1W_{12}(N_1;N_{\mathrm{tot}})$. Conversely, in the case where atoms transit from state 2 to state 1, we have $\nu(N_1)=N_2W_{21}(N_2;N_{\mathrm{tot}})$. If we assume that the total number of atoms is conserved, we find $\nu(N_1)=(N_{\mathrm{tot}}-N_{1})W_{21}(N_{\mathrm{tot}}-N_{1};N_{\mathrm{tot}})$. The stationary solution of Eq.~(\ref{derivalprodistribution}) then becomes,
\begin{eqnarray}
\label{prodistribution}
P^{(st)}_1(N_1)=P^{(st)}_1(0)\prod_{N=1}^{N_1}\frac{\nu(N-1)}{\mu(N)}.
\end{eqnarray}
For the PDDO one has $\alpha^{12}=\alpha^{21}$ and $\beta^{12}=\beta^{21}$, and substituting  $\mu(N_1)=N_1W_{12}(N_1;N_{\mathrm{tot}})$ and $\nu(N_1)=(N_{\mathrm{tot}}-N_{1})W_{21}(N_{\mathrm{tot}}-N_{1};N_{\mathrm{tot}})$ into Eq.~(\ref{prodistribution}), $P^{(st)}_1(N_1)$ is given by,
\begin{eqnarray}
\label{prodis}
P^{(st)}_1(N_1)\equiv P(x)\approx \tilde{Z}^{-1}\exp[-N_{\mathrm{tot}}(x^4-6\theta x^2)/12],~~\theta=\alpha N_{\mathrm{tot}}-1~(>0,<0),
\end{eqnarray}
where  $\alpha=\alpha^{12}=\alpha^{21}$, the quasi-continuous variable $x=(N_2-N_1)/N_{\mathrm{tot}}$ for large $N_{\mathrm{tot}}$ and for $|x|\ll1$, and $\tilde{Z}$ is the normalization constant.
Remarkably, the system can be equivalently described by the general hamiltonian corresponding to the Landau free energy, $\mathfrak{L}\equiv N_{\mathrm{tot}}(x^4-6\theta x^2)/12$, and thus has the standard form of the mean-field probability distribution near the symmetry-breaking transition. The order parameter of the mean-field transition is defined by $\eta=\langle x\rangle=\int^\infty_{-\infty} dx xP(x)/\int^\infty_{-\infty}dxP(x)$, and therefore one can account for the spontaneous symmetry breaking of the system using $\mathfrak{L}$. In the following section, we discuss the details.

In contrast to the PDDO, all of the $\alpha^{nm}$ and $\beta^{nm}~(n,m=1,2)$ for the RDDO are different and not same, and consequently there is no general hamiltonian like Eq.~(\ref{prodis}) for the quasi-continuous variable $x$. Nonetheless, interestingly, the amplitudes for the RDDO  (i.e., the large- and small-amplitude vibrational motion) can be analogous to the two phases of liquid and gas, respectively, as indicated by the similar observations of the sudden and sharp change of the population of the bistable state, resulting from the enhanced fluctuations near the driving frequency $\omega_{eq}$. Hence one needs to obtain the average of the population in each state for this oscillator, which can be done by calculating the time derivative of the average of the population in the state 1, $\partial_t\left<N_1\right>$.
It is given by,
\begin{eqnarray}
\partial_t\left<N_1\right>=\partial_t\left[\sum^{N_{\mathrm{tot}}}_{N_1=0}N_1P_1(N_1)\right],
\end{eqnarray}
where $\sum^{N_{\mathrm{tot}}}_{N_1=0}P_1(N_1)=1$.
Because $N_{\mathrm{tot}}\gg1$, $\partial_t\left<N_1\right>$ is simplified as,
\begin{eqnarray}
&&\partial_t\left<N_1\right>=\sum^{\infty}_{N_1=0}\left[\nu(N_1)-\mu(N_1)\right]P_1(N_1).
\end{eqnarray}
Therefore, in the steady-state solution, $\partial_t\left<N_1\right>=0$, and thus the following condition is given,
\begin{eqnarray}
&&\nu(N_1)=\mu(N_1)\rightarrow N_{2}W_{21}(N_{2};N_{\mathrm{tot}})=N_1W_{12}(N_1;N_{\mathrm{tot}}).
\end{eqnarray}
From the above formulas, one can obtain the ratio of $N_1$ to $N_2$ as,
\begin{eqnarray}\label{population}
&&\frac{N_1}{N_2}=\exp\left[-(\Delta R^{(0)}+\Delta R^{(1)})/D\right],
\end{eqnarray}
where
\begin{eqnarray}
\label{activationdiff}
&&\Delta R^{(0)}=R^{(0)}_2-R^{(0)}_1,\nonumber\\
&&\Delta R^{(1)}=D[2(\alpha^{21}N_2-\alpha^{12}N_1)\nonumber\\
&&~~~~~~~~~~~~-(\alpha^{21}-\alpha^{12}+\beta^{21}-\beta^{12})N_{\mathrm{tot}}].
\end{eqnarray}
 Here $\Delta R^{(0)}$ is the activation-energy difference between the two states in the noninteracting case, and $\Delta R^{(1)}$ is the activation energy difference between them, induced by the atom-atom interaction, depending on the total number of atoms.
Equation~(\ref{population}) can be  simplified further in terms of $\mathcal{P}_1(\equiv\frac{N_1}{N_{\mathrm{tot}}})$ as follows,
\begin{eqnarray}
&&\mathcal{P}_1=\frac{1}{1+\exp[\Delta R(\mathcal{P}_1)/D]},\label{population1}
\end{eqnarray}
where $\Delta R(\mathcal{P}_1)\equiv\Delta R^{(0)}+\Delta R^{(1)}$ and $\mathcal{P}_1+\mathcal{P}_2=1$.
Uding the above equation, one can the  calculate numerically the density of the population, $\mathcal{P}_{1(2)}$.

\subsubsection{Comparison of the interatomic interaction effect for PDDO and RDDO}
Until now, we have taken a closer theoretical look at the interatomic interaction effect for the cold atomic PDDO and RDDO, but nonetheless it is rather tedious and not easy to  understand intuitively their apparent distinctions.
\begin{figure}[thb] \centering
\includegraphics[scale=0.45]{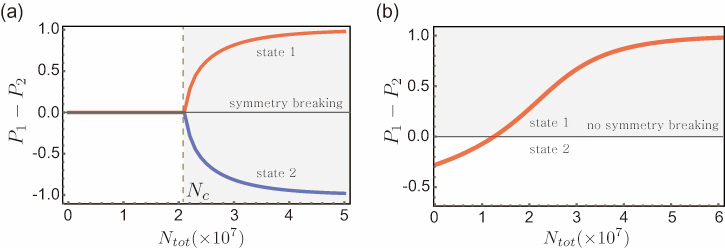}
\caption[figure13]{\label{collectivebehavior}\small Occupied population difference between the two bistable states in (a) PDDO and (b) RDDO due to the cumulative effects of the interaction of the switching atom with the atoms in the clouds. In (a), symmetry breaking occurs with the increase of $N_{\mathrm{tot}}$, occupying either of the two states (red or blue line), while in (b), the population of the state~1 only increases with $N_{\mathrm{tot}}$ due to the one-way bias effect of the interaction.}
\end{figure}
Figure~\ref{collectivebehavior} shows the dependence of the occupied population difference between the two stable states on the total number of atoms. Figure~\ref{collectivebehavior}(a) represents the spontaneous symmetry breaking;  all the atoms in the two stable states of PDDO spontaneously congregate in either side (red or blue lines) of the two stable states above the critical point, due to the symmetric contribution of the interatomic interaction between the two bistable states. On the other hand, in the case of RDDO, atoms are  populated only in one stable state corresponding to the large-amplitude state (stable state 1), without the property of symmetry breaking, due to the effect of the asymmetric interaction between them. For this reason, the spontaneous time-symmetry breaking occurs only in the PDDO due to the symmetric interaction mechanism, while the effect of the asymmetric interaction in the RDDO only causes the shift of the kinetic phase transition boundary where the population between the two stable states is equal~(see Sec.~5). The distinct roles of the light-induced interaction in the two nonlinear dynamics produce the different types of nonequilibrium phase transitions, such as the Ideal mean-field transition in the PDDO analogous to the Ising-type phase transition and the kinetic phase transition in the RDDO analogous to the discontinuous phase transition of the liquid-gas phase system.

\section{Spontaneous time-symmetry breaking}
\label{sec4}
When the Hamiltonian governing the dynamics of a system has a certain symmetry, its ground state can
have a lower symmetry and most of the phase transitions are accompanied by spontaneous breaking of this symmetry \cite{Landau:StatMech,Chaikin1995,Landau1937}.  In thermal equilibrium, in particular, phase transitions usually occur along with spontaneous breaking of the spatial symmetries. For example, the rotational symmetry of the Ising spins is broken under the ferromagnetic transition. When a liquid is solidified to a crystal, its symmetry under the \textit{continuous} spatial translation is broken and the crystal exhibits a new form of the \textit{discrete} spatial symmetry. Because the ground states in thermal equilibrium are stationary in time, the time symmetry is usually preserved. However, for nonequilibrium systems, the symmetry in time can  be also broken \cite{Khemani2016,Zhang2016,VonKeyserlingk2016,Yao2016}.

This section deals with spontaneous breaking of the symmetry associated with the discrete time translation in a periodically driven interacting system. Even though it happens out of equilibrium, one may employ the arguments  widely used for the phase transitions in thermal-equilibrium systems. Generally, the spontaneous symmetry breaking (SSB) occurs  by the interplay between the competing effects of thermal fluctuations and interactions. The relevant critical exponents, which are used to classify the class of the critical phenomena in thermal equilibrium, are also experimentally observed (Sec.~\ref{sec4.1}) and are shown in good agreement with the theoretical calculations (Sec.~\ref{sec4.2}).

\subsection{Experimental observations of critical properties}
\label{sec4.1}

\begin{figure}[h]\centering
\includegraphics[scale=0.5]{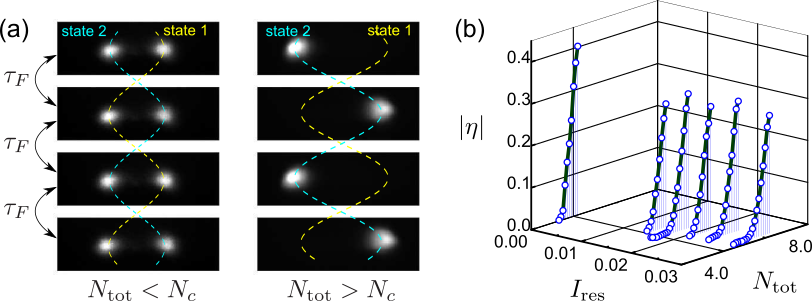}
\caption[figure14]{\label{fig:fig1}\small (a) CCD images taken for each modulation period $\tau_F$ at the positions where two states are spatially separated most. The right (left) panel is for the total number $N_{\rm tot}$ of atoms larger (smaller) than $N_c$. Trajectories of each vibrating state are depicted as the dashed curves. (b) Plot of the order parameter as a function of the intensity $I_{\rm res}$ of the resonant light and the total number $N_{\rm tot}$ of atoms.}
\end{figure}

The system of our interest consists of cold atoms in the PDDO (Eq.~\ref{nonlineareq}(a)), where two identical period-two states develop out of phase with the period $2\tau_F$, and atoms can populate either of the two states depending on their initial conditions. Because thermal fluctuations induce switching of atoms between the two states, they are equally populated regardless of their initial states, so that the system has the same symmetry as that of the equation of motion, as observed on the left panel in Fig.~\ref{fig:fig1}(a). This reminds one of the paramagnetic states of the Ising system where spins are randomly distributed among two spin states due to the thermally induced spin flips. Following this analogy with the Ising model, one may further expect the relevant symmetry can be spontaneously broken leading to a phase transition if one can realize the condition that the interaction dominates the thermal fluctuations.

For the PDDO, the interactions between atoms come from the shadow effect \cite{Sesko1991,Walker1990} as described in Sec.~\ref{subsec:interaction}. Here we want to emphasize again its two important features. First, this force is much weaker than the overall trapping force, but can still affect the switching dynamics of single atoms because the switching rate is exponentially sensitive to the strength of this force as shown in Sec.~\ref{Switching rate modification}. The second important aspect of this force is that it acts like an accumulated mean-field force as described in Sec.~\ref{subsec:interaction} and Fig.~\ref{fig:shadow}. Because the oscillation frequency of atoms ($\sim\omega_0$) inside the atomic cloud as well as their decay rate $\sim\gamma$ are much higher than the switching rate ($\sim 1$ s$^{-1}$), the intra-cloud density fluctuations decay quickly so that they are not correlated with the inter-cloud density fluctuations that characterize the atomic switching rate. Therefore, the dynamics of single atoms is independent of the spatial fluctuations and thus not affected by the spatial distribution of atoms, but determined solely by the averaged forces from the other atoms, or the number of atoms in each state, which is the essence of the mean-filed approximation. This is similar to the mean-field analysis of the Ising model, where a single spin interacts with the other spins with the same strength of interaction, so its dynamics is determined by the number of spins in each of spin state. This mean-field feature will be elucidated by the experimental and theoretical observations of critical exponents in the following sections.

We now make a further investigation of the experimental observations of SSB in the periodically driven trapped atoms. The symmetry under a discrete time translation by $\tau_F$ is preserved either at high thermal fluctuations or at low number of trapped atoms that is directly related to the strength of the mean-field interactions as was described in the previous paragraph. Figure~\ref{fig:fig1}(a) shows the snapshots of atoms at every $\tau_F$ for the spatial phases where the two states are most separated. When the total number of atoms, $N_{\rm tot}$, is smaller than a certain critical number of atoms $N_c$   for a  given temperature $T$, the system is invariant under each discrete-time translation by $\tau_F$. If $N_{\rm tot}$ is bigger than $N_c$, on the other hand, the atoms become populated preferentially in one of the two states and thus the system has a symmetry under the discrete time translation by $2\tau_F$, not by $\tau_F$. From these observations one can define the order parameter $\eta$ as the normalized difference of the number of atoms in each state, while the control parameter $\theta$ is defined as the total number of atoms with respect to its critical value,
\begin{equation}
\eta\equiv\frac{\left<N_2-N_1\right>}{N_{\rm tot}},\quad\theta\equiv\frac{N_{\rm tot}-N_c}{N_c}.
\end{equation}
As another controlling knob, one can also tune the amount of thermal fluctuations by adjusting the intensity $I_{\rm res}$ of a resonant light illuminated on the atoms. Dependences of the order parameter on $I_{\rm res}$ and $N_{\rm tot}$ are presented in Fig.~\ref{fig:fig1}(b). Here, we treat the total number of atoms as a control parameter.
\begin{figure}[h]\centering
\includegraphics[scale=0.5]{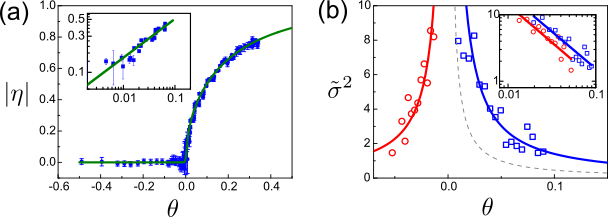}
\caption[figure15]{\label{fig:fig2}\small (a) Plot of the order parameter $\left|\eta\right|$ as a function of the normalized total number of atoms $\theta$. The solid curve is derived  from Eq.~(\ref{eq:tanh}). The log-log plot in the inset shows the power-law dependence of $\left|\eta\right|$ on $\theta$ for $\theta>0$ in the vicinity of $\theta=0$. The solid line represents the linear fit giving the critical exponent $\beta=0.51(1)$. (b) Plot of the variance $\tilde \sigma^2=10^3\sigma^2$ for the order parameter as a function of $\theta$. The curves are the fitted results obtained by  Eq.~(\ref{eq:def_exponents_theta}). The inset is a log-log plot that provides the critical exponents $\gamma_{\pm}$, where the critical exponents in the symmetric and the broken-symmetry phases are, respectively, $\gamma_-=1.04\pm0.21$ and $\gamma_+ = 1.11\pm 0.13$. (Data taken from Ref.~\cite{M Heo PRE 2010})}
\end{figure}
As for the thermal equilibrium systems, we characterize this SSB by observing the asymptotic behaviors of the order parameter and its fluctuations near the critical point. It has been known that they show the power-law dependences on the distance from the critical points such that,
\begin{eqnarray}
&&|\eta|\propto\theta^{\beta}\qquad\mbox{for }\theta>0,\label{eq:def_exponents_beta}\\
&&\sigma^2\equiv\langle x^2\rangle-\langle x \rangle^2\propto\theta^{-\gamma_{+(-)}}\qquad\mbox{for }\theta>0\;(\theta<0). \label{eq:def_exponents_theta}
\end{eqnarray}
Here $x \equiv (N_2-N_1)/N_{\rm tot}$ so that $\eta = \left<x\right>$.
Notice that the critical exponents $\beta$ and $\gamma_{+(-)}$ determine the class of  critical phenomena. Figure~\ref{fig:fig2} shows the experimental results, $\beta=0.51(1)$, $\gamma_+=1.11\pm0.13$ and $\gamma_-=1.04\pm0.21$, which agree well with the ideal mean-field predictions. These values imply that the SSB in the present system can be thought of as the mean-field phase transition.

\subsection{Theoretical descriptions}\label{sec4.2}
The steady-state probability $P_1(N_1,t)$ that the cloud 1 has $N_1$ atoms at time $t$ is given by Eq.~(\ref{prodis}). The distribution $P_1^{\rm st}(N_1)$ versus $\theta$ has one sharp peak at $x=0$ for $\theta<0$, while two sharp peaks at nonzero value of $x$ for $\theta>0$, which indicates that either of the two clouds is more populated. The value of the order parameter $\eta$ for a given control parameter $\theta$ can be calculated directly from Eq.~(\ref{prodis}) by finding $x_0$, which results in the extrema of $P_1(x, \theta)$. Since $\langle x\rangle=x_0=\eta$, the order parameter satisfies,
\begin{equation}\label{eq:tanh}
\eta=\tanh[(\theta+1)\eta].
\end{equation}
We now turn to the critical behaviors of the order parameter and its related quantities near the critical point $\theta=0$. It has been well understood that in thermal equilibrium those quantities follow the power-law dependence close to the critical point. Although the system we are interested in is out of equilibrium, we find the similar kinds of the power-law behaviors and their critical exponents. In addition, the values of these exponents are found to be identical to those from the ideal mean-field theory, which justfies our previous arguments on the interactions. Notice that from Eq.~(\ref{prodis}) or Eq.~(\ref{eq:tanh}), the order parameter increases with the control parameter near the critical point such that,
\begin{eqnarray}
\label{eq:critical_order_parameter}
\eta =\pm (3\theta)^{1/2}\quad {\rm for } \; 0<\theta \ll 1.
\end{eqnarray}
This shows that the closer to the critical point, the more the order parameter fluctuates. For $\left|\theta\right|\ll1$, the variance of the order parameter, $\sigma^2=\langle x^2\rangle-\eta^2$ that characterizes the fluctuations, is obtained by,
\begin{eqnarray}
\label{eq:population_fluct_symmetric_no_field}
&\sigma^2 = (N_c|\theta|)^{-1}\quad {\rm for}\; \theta<0, \nonumber\\
&\sigma^2=(2N_c|\theta|)^{-1}\quad {\rm for}\; \theta> 0.
\end{eqnarray}
Interestingly, Eq.~(\ref{eq:critical_order_parameter}) and Eq.~(\ref{eq:population_fluct_symmetric_no_field}) are very similar to those results derived by the conventional ideal mean-field theory.
\subsection{Response to the symmetry breaking field}
\label{Response to the symmetry breaking field}
When a weak modulation at the  frequency of $\omega_F/2+\Omega$ is added to the strong parametric modulation, the degeneracy of the two vibrational states is expected to be lifted. Moreover, the modulation changes periodically in time under $\Omega\neq0$ but remain static under $\Omega=0$ in the rotating frame. The situation is analogous to the well known magnetic spin system under an external magnetic field, and it provides a useful tool for exploring the novel  properties, such as the susceptibility, the dynamic response and the related critical phenomena. Experimentally the additional modulation is realized by adding a small modulation signal at the frequency $\omega_F/2+\Omega$ to the parametric modulation signal, where its amplitude is up to one hundredth of the parametric modulation amplitude.

\begin{figure}[h]\centering
\includegraphics[scale=0.5]{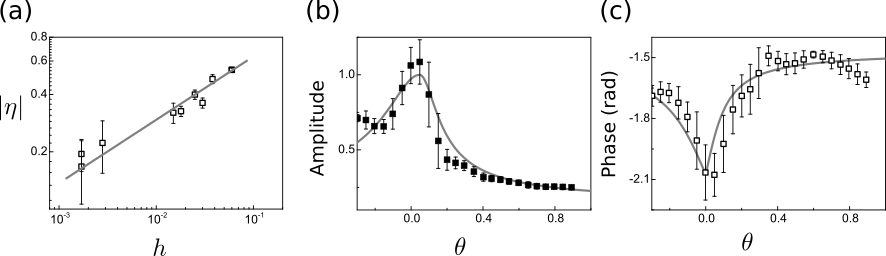}
\caption[figure16]{\label{fig:fig3}\small Response to the effective bias field. (a) The order parameter at the criticality ($\theta =0$) as a function of the amplitude $h$ of the additional modulation at the frequency  $\omega_F/ 2$. Here $h$ is scaled with respect to the strong modulation amplitude. The solid line shows the behavior $|\eta|\propto h^{1/\delta}$ at $\delta= 3$; experimentally, we find $\delta=3.0\pm0.8$.  The amplitude (b) and the phase (c) of oscillations of the order parameter $\eta$ are induced by an extra modulation at the frequency $\omega_F/2+\Omega$ for $\Omega= 0.1$~Hz. In (b), the amplitude is scaled by the value for $\theta=0$. The solid curves show the theory, Eq.~(\ref{eq:susc_omega_explicit}), with the $N_{\mathrm{tot}}\rightarrow0$ switching rate used as a fitting parameter. The error bars denote the standard deviations of 100 measurements. (Data taken from Ref.~\cite{M Heo PRE 2010})}
\end{figure}

As already mentioned, the discrete time translational symmetry of the system can be broken when one adds an additive periodic force ${\bf h}(t)=\hat z h \cos(\omega_Ft/2+\phi_h)$ $(h>0, \Omega=0)$ \cite{Dykman PRE 2006,Kim2010}. This force acts like a DC bias field in the  frame rotating at the frequency $\omega_F/2$. For small $\bf h$, the activation energy $R_{1,2}$ is modified such that,
\begin{eqnarray}
\label{eq:activ_energy_with_lifted_symmetry}
&&R_n=R^{(0)} + R_n^{(1)}+ R_n^{(h)}, \nonumber\\
&&R_n^{(h)}=-\int dt {\bm \lambda}_n^{\rm (opt)}(t){\bf h}(t)=\bar h \cos \phi_n,
\end{eqnarray}
where $\bar h\propto \left|\bf h\right|$ and $\phi_{1,2}$ is in general linear in $\phi_h$. This small change in the activation energy modifies substantially the switching rates $W_{1,2}$ and in turn, the stationary population distribution in Eq.~(\ref{prodis}) is then changed as,
\begin{eqnarray}
\label{eq:field_modified_G}
P_1^{\rm st}(N_1)\to P_1^{\rm st}(N_1)\exp\left(N_{\rm tot}x h_{12}\right),~h_{12}=-\bar h \cos\phi_1 /D.
\end{eqnarray}
Therefore the maximum of the probability satisfies the equation $\theta x_0-\frac{1}{3}x_0^3+h_{12}=0$. Moreover, the order parameter $\eta$ at the critical point $\theta=0$ follows,
\begin{eqnarray}
\eta=x_0\propto\bar h^{1/3},
\end{eqnarray}
which is in good agreement with the experimental results in Fig.~\ref{fig:fig3}(a).

One can also induce an effective AC bias field by detuning the frequency of the additive periodic force by a small amount $\Omega$, ${\bf h}(t)=\hat z h \cos(\omega_F/2+\Omega)t$, where $|\Omega|\ll \gamma\ll \omega_F$. This produces a modulation of the activation energy in Eq.~(\ref{eq:activ_energy_with_lifted_symmetry}) giving,
\begin{eqnarray}
\label{eq:periodic_modulation}
R_1^{(h)}\equiv R_1^{(h)}(t)=-R_2^{(h)}(t)= \bar h\cos(\Omega t).
\end{eqnarray}
Here one can linearize the switching rate $W_{nm}$ with respect to $\bar h$. When   the generalized frequency-dependent susceptibility $\Xi(\Omega)$ is defined in the form,
\[\delta\eta(t)\equiv \langle x \rangle-x_0= -\frac{1}{2D}\left[\Xi(\Omega)\bar h\exp(-i\Omega t) +{\rm c.c.}\right],\]
where $x_0$ is the position of the maximum of the distribution for $\bar h=0$,
one can then obtain \cite{Heo2010},
\begin{eqnarray}
\label{eq:susc_omega_explicit}
&&\Xi(\Omega)= 2\tilde W/(2|\theta|\tilde W -i\Omega), \qquad \theta < 0,\nonumber\\
&&\Xi(\Omega)= 2\tilde W/(4\theta\tilde W -i\Omega), ~~\qquad \theta > 0,
\end{eqnarray}
where $\tilde W=W^{(0)}\exp(\beta N_{\mathrm{tot}})$ and $W^{(0)}\equiv W_{12}^{(0)}=W_{21}^{(0)}$.

%In particular, the dynamic response of the parametrically modulated MOT system under oscillating bias field can be simply described by the mean-field theory.
\begin{figure}[h]  \centering
\includegraphics[scale=0.45]{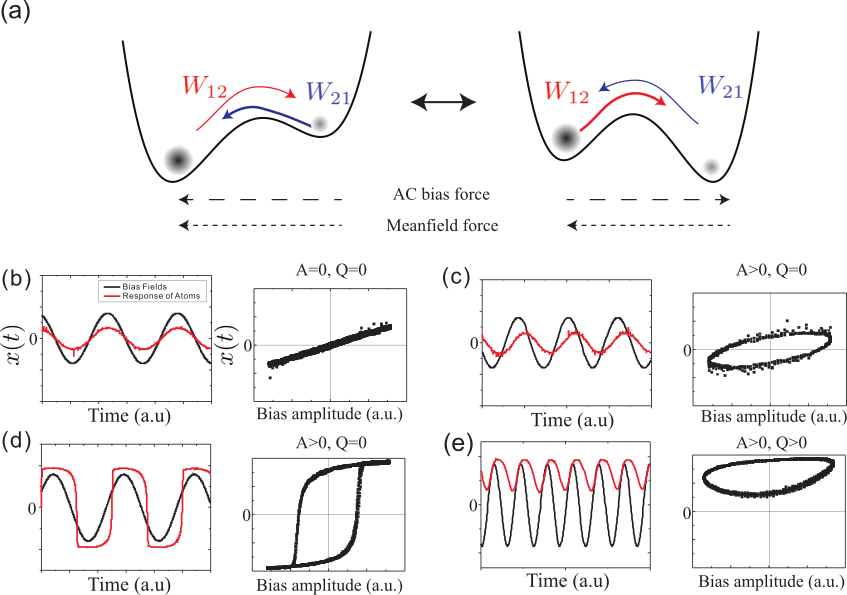}
\caption[figure17]{\label{fig:response}\small  Dynamic response of the normalized population difference $\langle x(t)\rangle$ to the effective AC bias fields. This has an analogy with the double-well potential driven by a periodic force as seen in (a). The AC bias field tilts the double well periodically but the mean-field attractive force is always directed toward the more populated state. This causes the barrier difference on the right panel to be smaller than that on the left one. On the left panels from (b) to (e), the measured $\langle x(t)\rangle$ (red solid curves) are shown  for different values of the frequency $\Omega $ of the AC bias field (black sinusoidal curves) and the total number of atoms $N_{\rm tot}$.  The corresponding hysteresis curves are presented on the right panels of (b) to (e).  The hysteresis loop area $\text{A}$ as well as the dynamic order parameter $\text{Q}$ are also indicated. Each data is obtained when (b) $\Omega \ll1$ Hz and $N_{\rm tot}<N_c$, (c) $\Omega\sim1$ Hz and $N_{\rm tot}<N_c$, (d) $\Omega\ll1$ Hz and $N_{\rm tot}>N_c$ and (e) $\Omega\sim1$ Hz and $N_{\rm tot}>N_c$. }
\end{figure}

In the following, we now discuss the dynamic responses to the AC bias fields for various amplitude $\bar h$ and frequency $\Omega$. Using the activation energy of Eqs.~(\ref{eq:activ_energy_with_lifted_symmetry}) and (\ref{eq:periodic_modulation}), the noise-induced transition rates with the interaction between atoms included and the additional  AC bias field can be rewritten as,
\begin{equation}
\label{switching_rate_bias_interaction}
W_{12}=\tilde W \exp\left[(\theta+1) x+\tilde h(t)\right],~~W_{21}=\tilde W \exp\left[-(\theta+1) x-\tilde h(t)\right],\quad\tilde h(t)=\bar h \cos\Omega t .
\end{equation}
%By writing down a rate equation of $dN_1/dt$ and $dN_2/dt$ using the Eq.~(\ref{switching_rate_bias_interaction}), we can get the deterministic equation of $x$ under the oscillating bias field as follows,
%\begin{eqnarray}
%\label{dynamics_eq}
%\frac{1}{2\tilde W}\frac{dx}{dt}=-x\cosh[(\theta+1)x+\bar h \cos\Omega t]+\sinh[(\theta+1)x+\bar h \cos\Omega t].
%\end{eqnarray}
%We may be able to solve this equation numerically to obtain the system response to the AC bias field. But let us provide with a qualitative description. 
The dependency on the periodic bias field has an analogy with the double-well system, or  the two-state system,  driven by a periodic force as depicted in Fig.~\ref{fig:response}(a).  Here the quantities $x$ and $\theta$ are related to the mean-field interaction term that acts as the bias field proportional to the atom number difference $  N_2-N_1  \propto x$. Because the equation of motion remains invariant under $t \rightarrow t+2\pi/\Omega$, the system response also becomes periodic with the same frequency $\Omega$ or an integer multiple of it. To facilitate further discussion, one can introduce the new dynamic parameters; the area of the hysteresis loop $\text{A}$ and the period-averaged order parameter $\text{Q}$. These two dynamic quantities are defined by the values averaged over a complete period as follows~\cite{ChakrabartiRMP1999},
\begin{eqnarray}
&&\text{A} \equiv \oint x(t) d\tilde{h}, \\
&&\text{Q} \equiv \frac{\Omega}{2\pi}\oint x(t)dt.
\end{eqnarray}

 The hysteresis loop area $\text{A}$ is originated from the phase delay between $x(t)$ and $h(t)$, when the frequency of the bias field is large enough so that the system  cannot follow in response. The hysteresis loop area $A$ vanishes only when both $\Omega$ and $N_{\rm tot}$ are small enough for the system to  follow quickly the bias field as in Fig.~\ref{fig:response}(b). The period-averaged order parameter $\text{Q}$ vanishes when the frequency $\Omega$ of the bias field is low so that the period of the bias field is long compared with the system response time as seen in Fig.~\ref{fig:response}(b), (c) and (d).  When the bias-field period is shorter than the effective relaxation time of the system, the dynamically broken symmetric phase can arise spontaneously with the nonvanishing values of $\text{Q}$ as seen in Fig.~\ref{fig:response}(e). It is characterized by the change of the average value, $\langle \text{Q}\rangle=0\rightarrow\langle \text{Q}\rangle\neq0$, although the external field does not induce any symmetry breaking field over the period. For this reason, one can call $\langle \text{Q}\rangle$ the dynamic order parameter. For the study of the dynamic responses, one needs to control both the period of the bias field $\tau_{\rm bias}$ and the system relaxation time $\tau_{\rm r}$. $\tau_{\rm bias}$ can be easily tuned by the bias oscillation frequency $\Omega$. On the other hand, $\tau_{\rm r}$ is related to the switching rate $W_{12(21)}$ in Eq.~(\ref{switching_rate_bias_interaction}), which is affected by  the strength of the bias field $\bar h$ and the interaction as well as the the thermal fluctuations. One can thus control $\tau_{\rm r}$ by changing the total number of atoms and the amplitude of the bias field.

 When the bias varies sufficiently slowly (i.e., very long $\tau_{\rm ext}$) for $N_{\rm tot}<N_c$, the system has enough time to follow the instantaneous value of the bias. As a result, the order parameter $x$ does not depend on the past trajectory and hence there is no hysteresis as shown in Fig.~\ref{fig:response}(a). With the increase of the frequency of the bias, however, the delay of the system response is developed so that the inversely symmetric hysteresis loops appear as shown in Fig.~\ref{fig:response}(b). For $N_{\rm tot}>N_c$, the hysteresis similar to Fig.~\ref{fig:response}(c) always exists even at the very low $\Omega$ because of the metastability of the system. Notice that for $N_{\rm tot}>N_c$, the hysteresis develops from the triple-valued stationary solutions of $x$ (Eq.~(\ref{prodis})) with $\bar h=0$, while for $N_{\rm tot}<N_c$ the hysteresis results from the single-valued one.  Remarkably, when the frequency of the bias field increases above the certain value for $N_{\rm tot}>N_c$, the hysteresis loop then becomes asymmetric around the origin as shown in Fig.~\ref{fig:response}(d) so that $\langle \text{Q}\rangle > 0$ is produced.

\subsection{Observation of dynamic critical behavior in the PDDO}
For the globally coupled PDDO that is considered here, one can observe  two different kinds of dynamic critical behaviors, one at the critical point of the SSB originating from the collective behavior and the other at the bifurcation point of the PDDO ($\omega_1$ in Fig.~\ref{fig:vibrational motion}(a)), which is basically of the single-particle behavior. In Sec.~\ref{sec:critical clowing down}, we discuss the critical slowing down at the critical point $N_{\rm tot}=N_c$ that occurs as one sweeps the total number of atoms $N_{\mathrm{tot}}$, which is due to the interplay between the noise-induced switching and the interatomic interaction as also observed in other interacting many-particle systems~\cite{P.C. Hohenberg1977}. In Sec.~\ref{sec:relaxation of unstable state}, we investigate the relaxation behavior  in the vicinity of a bifurcation point for $N_{\mathrm{tot}}\ll N_c$, which ensures the single-particle dynamics, and observe the universal features as also found in other systems~\cite{Krivoglaz1980,JETF 1979,M. I. Dykman 5902,Bryant 1986}. 

\subsubsection{Critical slowing down}\label{sec:critical clowing down}
\begin{figure}[thb] \centering
\includegraphics[scale=0.4]{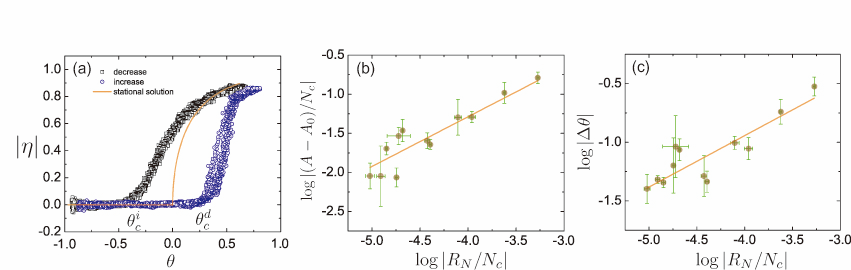}
\caption[figure18]{\label{hysteresis}\small Hysteresis curve and scaling exponent. (a) Measured hysteresis loops for the sweeping rate $1.87 \times 10^6$  s$^{-1}$. Log--log plot of (b) the hysteresis area $\mathcal{A}$ and (c) the hysteresis width $\Delta \theta$ versus the number sweeping rate $R_N$. The solid red lines in (b) and (c) are the logarithmic fits of Eqs.~(\ref{1}) and (\ref{16}), respectively. (Data taken from Ref.~\cite{wanheepre2016})}
\end{figure}

It has been known that the correlation length diverges near the critical point (e.g., critical temperature), and the region of the system, which represents the fluctuations near the equilibrium state, gets larger and larger. Namely, it takes longer and longer for the system to relax independent of the specific equilibration mechanisms. The phenomenon is called the \textit{critical slowing down}.
As is discussed in the previous section, the time-translational symmetry breaking displays the ideal mean-field transition, indicating the second-order phase transition or the continuous phase transition. Interestingly, as shown in Fig.~\ref{hysteresis}(a), when the total number of atom sweeps across the critical point ($N_c$) maintaining the uniform sweep speed, one can show that the system undergoes the critical slowing down~\cite{wanheepre2016}.

Although both the cold atom system and the magnetic spin system exhibit the similar critical slowing down phenomenon, the physical origin between the two systems is very different in the microscopic point of view. In the case of the resonantly driven cold atom system, there is no correlation length defined because it is effectively the zero-dimensional system (theoretical assumption) and is in nonequilibrium where the detailed balance is not satisfied. Near the critical point $N_c$, the noise-induced switching rate virtually diminishes, and in this situation, the attractive interaction among atoms contributes dominantly to the atomic transition rate between the two stable states. Consequently, the symmetry breaking thereby takes place.

The hysteresis, a nonequilibrium phenomenon obtained typically as the temperature of the system is varied, is one of the most interesting topics that have been studied in various fields such as the molecular switching using the spin crossover~\cite{R. G. Miller 2014,R. Kulamczewski 2014,M. Shigeno 2016}, the temperature driven metal-insulator transition in the solid-state devices~\cite{S. Singh 2012, R. Xie 2011} and the antifreeze proteins in the bionic systems~\cite{Can 2013, C. P. Garnham 2011}. The phenomenon of thermal hysteresis was also reported in the mean-field model~\cite{F. Zhong 1994}, for which the closed hysteresis loop area ($\mathcal{A}$) scales with the change rate of temperature $R_T$ as,
\begin{equation}\label{1}
\mathcal{A}=\mathcal{A}_0+b R_T^{\tilde\alpha},
\end{equation}
where $\tilde\alpha$ is the scaling exponent of the hysteresis, and $\mathcal{A}_0$ and $b$ are constants. It is known that $\tilde\alpha$ approaches the value 2/3, which is universal for both the mean-field and the field theoretical models~\cite{F. Zhong 1994, J. X. Zhang 1996}. Likewise, the spontaneous time-symmetry breaking transition in the parametrically excited atomic system displays the similar hysteretic behavior depending on the sweeping rate of the total number of atoms $R_N$ instead of $R_T$, and the universality class of the hysteresis is obtained through the scaling exponent of the hysteresis loops.

Figure~\ref{hysteresis}(a) shows the hysteresis curves as a function of the reduced atom number for the sweeping rate of $1.87 \times 10^6$  s$^{-1}$. The parameters $\theta_c^i$ and $\theta_c^d$ are the critical values of $\theta$ where the transition between $\eta=0$ and $|\eta|>0$ occurs for the increase and decrease of $\theta$, respectively. The shadow effect, i.e., the symmetry breaking interaction of the SSB transition, does not catch up with the variation in the number difference between the two clouds, because the time for the order parameter to relax toward the equilibrium state becomes longer and longer when the number of atoms approaches the critical number. Thus, when the total number of atoms is swept across the critical number at a uniform speed, the hysteresis loops can be obtained during the transition as shown in Fig.~\ref{hysteresis}(a). It can be observed that the area of the hysteresis loop decreases as the sweeping rate of the total number of atoms is decreased.

Figure~\ref{hysteresis}(b) presents the scaling behavior of the hysteresis loop area versus the sweeping rate of the total number of atoms on the log--log plot. Each point was obtained by averaging the experimental values more than three times and the constant term $\mathcal{A}_0$ was derived by a linear fitting of the hysteresis area $\mathcal{A}$ versus the sweeping rate $R_N$ plot. The scaling exponent $\tilde\alpha$ for our system was $0.64 \pm 0.04$, which is quite close to the value given by the mean-field theory~\cite{F. Zhong 1994}. Therefore, it is clearly seen that the hysteresis induced by the atom number sweeping in the parametrically resonant system exhibits the thermal hysteretic behavior.

We now define the hysteresis width by $\Delta \theta =\left| \theta_c^i -\theta_c^d \right|$, which is known to be described by the scaling law,
\begin{equation}\label{16}
\Delta \theta \cong R_N^{\tilde\beta},
\end{equation}
where $\tilde\beta$ is the scaling exponent~\cite{S. Yildiz 2004}. The scaling exponent can be obtained by fitting the hysteresis width data to the double logarithmic form of Eq.~(\ref{16}). The log--log plot of the hysteresis width $\Delta \theta$ versus the sweeping rate $R_N$ is shown in Fig.~\ref{hysteresis}(c). The scaling exponent $\tilde\beta$ of the hysteresis width was found to be $0.44 \pm 0.025$, which is very close to the value of 0.465 predicted by the kinetic Ising model~\cite{K. A. Takeuchi 2008}. Following the scaling theory of thermal hysteresis developed by Zhong $et$ $al$. in Ref.~\cite{F. Zhong 1994}, the scaling exponent of the hysteresis width should have the physical meaning of the resistance characteristics for the glass transition. To describe the resistance of the system, the scaling exponent $\tilde\beta$ should be compared with the exponent $\tilde\alpha$ reported in the mean-field model~\cite{F. Zhong 1994,J. X. Zhang 1996,Y-Z. Wang 2011}. The value obtained in our system was $\tilde\beta \approx  0.44 \pm 0.025$, which is slightly smaller than the scaling exponent $\tilde\alpha$ of the mean-field model but still in good qualitative agreement. This implies that our system corresponds to the thermal model with a rather low resistance.

\subsubsection{Relaxation of an unstable state}\label{sec:relaxation of unstable state}
The relaxation process of a macroscopic system that is initially prepared in an unstable state is an intriguing problem in nonequilibrium physics~\cite{Haken Synergetics}, as it is related to such research areas as the transient-laser radiation~\cite{Haake PRL 1978,Suzuki JSP 1977a,Arecchi PRL 1971,Arecchi PRA 1975}, the spinodal decomposition~\cite{Langer PRA 1975,Kawasaki PRA 1978}, the superfluorescence~\cite{Haake PRL 1979,Haake PRL 1980,Polder PRA 1979} and the hydrodynamic instabilities~\cite{Whitehead JFM 1969}. Extensive works in this field have been performed theoretically from various perspectives~\cite{Haake PRL 1978,Risken ZP 1967,Caroli JSP 1979,Arecchi PRL 1980,Suzuki JSP 1977b,Pasquale PRA 1982}.

One of the characteristics of the relaxation process from an unstable state is the scaling behavior, and the relaxation dynamics near the bifurcation point shows the scaling behavior depending on the nature of the initial state and the type of instability involved~\cite{geolmoonpre2011}. There have been many studies of relaxation dynamics in various nonlinear systems, such as the saddle-node bifurcation, the pitchfork bifurcation and the super-critical Hopf bifurcation~\cite{S. H. Strogatz 2001}. Here, we discuss  the relaxation dynamics near the sub-critical Hopf bifurcation point~\cite{Holzner 1987,Baugher 1989,Larger 2004} utilizing the parametrically modulated cold atoms in the magneto-optical trap (MOT).

As we have witnessed so far, the nonlinear and stochastic nature in the MOT system produces the parametric resonance and Duffing oscillation, the Hopf bifurcation~\cite{K. Kim 2004,KihwanPRA:Hopf} and the noise-induced transition between two attractors~\cite{K. Kim PRA 2005}.  In particular, in the parametrically excited atomic system, one can symmetrically realize two attractors in the phase space like the dynamic double well and an unstable state at the center in the phase space, and hence the system provides the opportunity to do research on the relaxation dynamics of an unstable state.
\begin{figure}[thb] \centering
\includegraphics[scale=0.6]{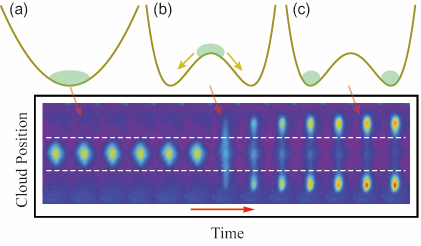}
\caption[figure19]{\label{Ch4_relaxation1}\small Experimental observation of the relaxation of an unstable state. The figure represents the typical 2D images detected at a time interval of 0.02 s. The oscillatory atomic motion is periodically observed at the specific timing when the two clouds are maximally apart. Initial preparation of the atomic cloud in the harmonic potential is shown in (a), and the subsequent location of the cloud on the unstable point in (b) and the fluctuation-induced decay of the atoms to the dynamical double wells in  (c). The typical experimental conditions are as follows: The magnetic field gradient $b$ is 0.11 T/m, the modulation amplitude of the trap laser $\epsilon$ is 0.8, and the saturation parameter $s_0$ and the detuning $\Delta$ are 0.2 and -2.55 $\Gamma$, respectively. Here $\Gamma$ is the decay rate of the excited state (=2$\pi\times$6.07 MHz).}
\end{figure}

\begin{figure}[thb] \centering
\includegraphics[scale=0.35]{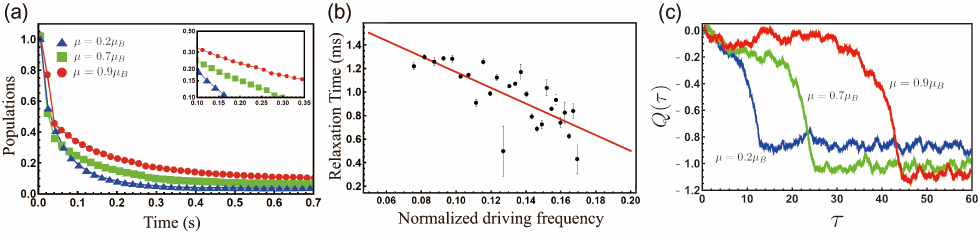}
\caption[figure20]{\label{Ch4_relaxation2}\small (a) Typical experimental decay of the atomic population at the center of atomic cloud, where the bifurcation point approaches from the blue to the red line in time sequence. (b) Scaling behavior of the relaxation process from the unstable state. The relaxation time diverges as the system approaches the bifurcation point, where the scaling exponent is -1.002 ($\pm$0.024) (log-log plot). The error bars represent the standard deviation for the fitting error. (c) Typical stochastic trajectories obtained by rhw numerical simulation for the initial condition of Y(0)=0 for different values of $\mu$. Note that  $D_0=0.5$ and $\zeta=1.275$. (Figures from Ref.~\cite{geolmoonpre2011})}
\end{figure}

To investigate the relaxation of an unstable state, the bifurcation point  is approached where the unstable state becomes stable.  While the driving frequency $\omega_F$ is changed  in the range between the super- and the sub-critical Hopf bifurcation, the distance ($\omega_B-\omega_F$) between the subcritical bifurcation point $\omega_B$ and the position of an unstable state $\omega_F$ should be adjustable~\cite{K. Kim 2004}.
Below the subcritical bifurcation point,  the unstable state is placed at the trap center in the phase space, while the stable states are located at  the points where the two clouds are maximally apart. To observe the escape process of the atomic population from the unstable state, one first has to locate the atoms at the trap center without any parametric excitation (Fig.~\ref{Ch4_relaxation1}(a)), after which one suddenly turns on the excitation (Fig.~\ref{Ch4_relaxation1}(b)). In such a situation, the trap center becomes unstable and the atoms begin to drift away from the center as time passes by, as show in Fig.~\ref{Ch4_relaxation1}(c). Figure~\ref{Ch4_relaxation2}(a)  displays experimentally the typical decay of the normalized atomic population density at the center of the atomic cloud for a different distance from the bifurcation point $\mu_B$. One can then obtain the relaxation time from the asymptotic exponential curves.

Figure~\ref{Ch4_relaxation2}(b) plots the relaxation time of the atomic population in an unstable state versus the normalized driving frequency experimentally rescaled by $(\omega_B-\omega_F)/\omega_B$, where  $\omega_B-\omega_F$ is proportional to $\mu_B-\mu$.  As one approaches the measured bifurcation point, $\omega_B~(=2\pi\times113.5\textrm{Hz})$,  the power-law behavior of the relaxation time is obtained. The measured scaling exponent is -1.002 ($\pm$0.024), which indicates that the system stays longer in the unstable state as the driving frequency $\omega_F$ moves closer to the bifurcation point $\omega_B$. The relaxation behavior displaying the scaling exponent can be theoretically examined by the Fokker-Planck equation describing the time evolution of the density distribution of the atomic cloud as below.

The stationary distribution of the atomic cloud in the phase space has the Gaussian form,
\begin{equation}
\label{distribution_real} \rho_0(z, v)=\frac{\gamma \omega_0}{2\pi
D}\exp[-\gamma(\omega_0^2 z^2 +v^2)/2D],
\end{equation}
and we focus on the relaxation process from the unstable equilibrium point $(P_{eq}, Q_{eq})=(0, 0)$ as mentioned in Sec.~\ref{the rotating frame}. Near the subcritical bifurcation point $\mu \approx \mu_B=\sqrt{1-1/\zeta^2}$, one direction of the motion of the system becomes slower than the other direction. Using the center manifold theorem~\cite{Nonlinear oscillator}, we can separate the fast variable $X$ and the slow variable $Y$ by applying an additional coordinate transformation, as follows~\cite{M. I. Dykman 5902},
\begin{equation}
X=P\cos\varphi-Q\sin\varphi,\quad Y=P\sin\varphi+Q\cos\varphi,
\end{equation}
where $\varphi =\frac{1}{2}\arcsin(1/\zeta)$. In this frame, Eq.~(\ref{rotating}) with $H_{\mathrm{sh}}=0$ has the form,
\begin{eqnarray}
\frac{dX}{d\tau}&=&-2\zeta^{-1}X+Y[\mu_B+\mu-(X^2+Y^2)]+\xi_X(\tau), \nonumber\\
\frac{dY}{d\tau}&=&X[\mu_B-\mu+(X^2+Y^2)]+\xi_Y(\tau).\label{eq_PQ}
\end{eqnarray}

The fast variable $X$ reaches its quasi-stationary value $X\approx\zeta\mu_B Y$ with the dimensionless relaxation time $\zeta/2$. On the other hand, the slow variable $Y$ has a relaxation time that diverges. Therefore, the fast variable $X$ follows adiabatically the slow variable $Y$, and thus one
can neglect the noise term in the equation of $X$. This finally allows the one-dimensional equation of motion for the slow variable $Y$, as given by,
\begin{equation}\label{eq_Q}
 \dot{Y}=\zeta \mu_B \delta\mu Y+\zeta^3 \mu_B Y^3+\xi(\tau),
\end{equation}
where $\delta\mu=\mu_B-\mu$, and $\xi(\tau)$ is the noise for the properly scaled intensity $D_0$. We consider that near the subcritical bifurcation point, the parameter $\delta\mu$ satisfies $0<\delta\mu\ll1$.

One can now calculate the many stochastic trajectories of Eq.~(\ref{eq_PQ}) as numerically  shown in Fig.~\ref{Ch4_relaxation2}(c).
The time evolution of the system can be described by the density distribution of the atomic cloud using the Fokker-Planck equation, which can be directly compared to the
experimental data. The corresponding Fokker-Planck equation of Eq.~(\ref{eq_Q}) is,
\begin{equation}
\frac{\partial{\rho}}{\partial{\tau}}=-\frac{\partial}{\partial{Y}}\left[(\zeta
\mu_B \delta\mu Y+\zeta^3 \mu_B
Y^3)\rho\right]+D_0\frac{\partial^2\rho}{\partial{Y^2}}.\label{FP}
\end{equation}
The initial condition is given by the distribution of Eq.~(\ref{distribution_real}), and transforming this initial distribution into the new frame gives,
\begin{equation}\label{distrbution_Q}
\rho(Y,\tau=0)=(2\pi \zeta D_0)^{-1/2}\exp(-Y^2/2\zeta D_0).
\end{equation}
The above equation can be easily solved in the restricted region where $Y^2\ll\delta\mu\zeta^{-2}$, implying that the nonlinear term is negligible. In this regime, the solution of Eq.~(\ref{FP}) with Eq.~(\ref{distrbution_Q}), or the time evolution of the distribution $\rho(Y=0, \tau)$ at the trap center, is initially in the nonexponential form. However, at a dimensionless time that satisfies $\exp(2\varepsilon\tau)\gg 1$ where $\varepsilon=\zeta \mu_B \delta\mu$, it shows the exponential decay behavior,
\begin{equation}\label{asymptotic_solution}
\rho(Y=0,\tau)\approx\sqrt{\frac{\varepsilon}{2\pi
D_0 (1+\zeta\varepsilon)}}\exp(-\varepsilon \tau).
\end{equation}
As a result, the asymptotic relaxation time $\tau_{r}$ demonstrates the scaling behavior that is inversely proportional to the distance from the bifurcation point,
\begin{equation}\label{decay_power_law}
\tau_{r}=1/\varepsilon\propto(\mu_B-\mu)^{-1},
\end{equation}
which also demonstrates that the scaling exponent is -1.

\section{Kinetic phase transition}
\label{sec5}
It has been well established that in thermodynamic systems, one of the intrinsic features of phase transition is the abrupt growth of fluctuations at the critical point or at the phase boundary~\cite{Goldenfeld}.
Even in the single-particle driven nonlinear system such as the RDDO, the prominent enhancement of fluctuations that results from the noise-induced switching between the coexisting states has been also observed  in the measured spectral density of fluctuations near the specific driving frequency. This is indicative of phase transition in nonequilibrium systems similar to the discontinuous phase transition of the gas and liquid phase in equilibrium, the so-called kinetic phase transition (KPT)~\cite{Dykman PRE 1994}.
However, it has been difficult to go beyond the single-particle behavior and investigate the many-body effects associated with the inter-particle interactions between the coexisting states. Therefore, here we try to devote our attention to the KPT in \textit{the weakly interacting ensemble system}~\cite{K. Kim PRA 2005}, which has not been studied despite many KPT works performed until now.
\subsection{Observation of the coexisting vibrating oscillators}
\begin{figure}[ht]\centering
\includegraphics[scale=0.45]{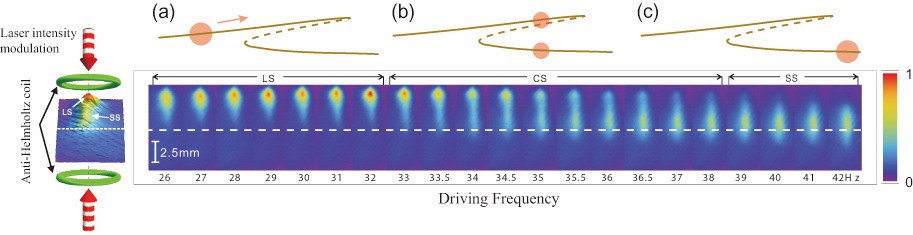}
\caption[figure21]{\label{Ch5_Fig1}\small Weakly interacting atomic RDDOs under the fluctuation-induced switching. The  fluctuations due to spontaneous emission of photons triggers the noise-induced switching between the two attractors. The images of the oscillatory atomic motion are captured at the regular intervals such that the large-amplitude state (LS) atomic cloud is maximally apart from the trap center (dashed horizontal line) whereas  the small-amplitude state (SS) cloud is not in the maximum vibrational amplitude state because of the phase difference between the LS and the SS.  The number of atoms is about $2.27\times10^7$ and the measured $\gamma$ and $\omega_{0}$ are 50.44 s$^{-1}$ and $2\pi\times28.26$ Hz, respectively. (Figures from Ref.~\cite{gmoon njp2013})}
\end{figure}

The RDDO exhibits two coexisting dynamical states (CS) as well as the large-amplitude state (LS) and small-amplitude state (SS), which have differing vibrational amplitude and phase in the specific range of driving frequency (Fig.~\ref{Ch5_Fig1}). If the atomic cloud is initially prepared in the LS  and one adiabatically sweeps the driving frequency towards the coexisting state region, the atoms occupied in the LS  start to transit to the SS  because the activation energy of the SS  is gradually increased, while that of the LS  becomes decreased. Therefore, depending on the driving frequency, the magnitude of the activation energy reverses and all the atoms will be occupied by the SS cloud. As opposed to the single-particle nonlinear oscillator, $10^6\thicksim10^7$ atoms in the RDDO ensemble system  are distributed over the dynamical states, and thereby one can see the cigar shaped atomic cloud. Furthermore, one can readily measure the steady-state populations occupied in each state without any statistical analysis of the realtime single particle trajectories.
\begin{figure}[ht]\centering
\includegraphics[scale=0.45]{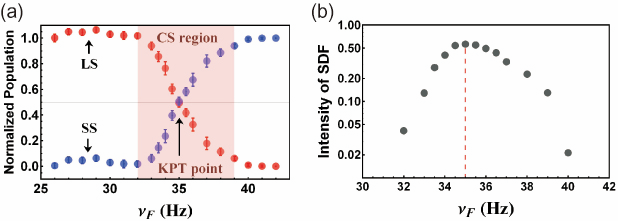}
\caption[figure22]{\label{Ch5_Fig2}\small (a) The change of atomic population in each state versus the driving frequency $\nu_F(=\omega_F/2\pi)$. The KPT occurs when the two populations are equal in the coexisting dynamical states (CS). (b) The measured intensity of the spectral density of fluctuation is maximized at the KPT point, leading to the onset of fluctuation enhancement linked to the transitions between the two states (semi-log scale). The error bars show the standard deviations. (Figures from Ref.~\cite{gmoon njp2013})}
\end{figure}

Figure~\ref{Ch5_Fig2}(a) shows the normalized atomic populations versus the driving frequency near the KPT point at  $\omega_F /2 \pi =35$ Hz. As shown, the two attractors are equally occupied at the KPT point because the activation energies of the two attractors are equal. Near the KPT point, the distinct feature of phase transition appears that  accompanies the large fluctuations associated with the transitions between the two states.

In order to observe the enhanced fluctuations in the case of the single particle system such as the micromechanical oscillator and the analog electronic circuit~\cite{L J Lapidus PRL 1999,J S Aldridge PRL 2005,R L Badzey Appl Phys Lett 2004,Dykman PRL 1990}, the spectral density of fluctuations $Q(\omega)$ (SDF) for the coordinate $z(t)$ of the oscillator was measured~\cite{Chan PRL 2006,Dykman PRL 1990},
\begin{eqnarray}
&&Q(\omega)=\frac{1}{\pi}\mathrm{Re}\int^\infty_0dt\exp(i\omega t)\mathcal{Q}(t),\\
&&\mathcal{Q}(t)=\lim_{T\rightarrow\infty}\frac{1}{2T}\int^T_{-T}d\tau[z(t+\tau)-\langle z(t+\tau) \rangle][z(\tau)-\langle z(\tau) \rangle].\nonumber
\end{eqnarray}
Here $Q(\omega)$ is separated into two contributions, $Q_i(\omega)$ associated with the small fluctuations in each state and $Q_{tr}(\omega)$ due to the noise-induced transitions between the two states, and as a result, $Q(\omega)=\sum_i\mathcal{P}_iQ_i(\omega)+Q_{tr}(\omega)$, where $\mathcal{P}_i$ is the density of  population in the $i$th state (Eq.~(\ref{population1})).
Interestingly $Q_{tr}(\omega)$ displays a very sharp and large peak due to noise-induced transition between the two dynamical states.
As shown in the above equation, to obtain the spectral density of fluctuations, one needs to record the trajectory in time. However, to confirm the large fluctuations near the KPT point in the system consisting of many particles, it is not possible to obtain the temporal traces of the single-particle trajectory. Instead of analyzing the temporal correlation of the single-particle trajectory~\cite{Chan PRL 2008,Chan PRL 2006}, the integrated spectral density of fluctuation $I$ (SDF) versus the distance from the KPT point can be obtained by the vibrational amplitude and the population of each state by using the relation~\cite{Dykman PRE 1994},
\begin{eqnarray}\label{Duffing}
I \equiv\int^\infty_{-\infty}d\omega Q_{tr}(\omega)=\frac{(z^l_{max}-z^s_{max})^2}{4}\mathcal{P}_1\mathcal{P}_2 ,
\end{eqnarray}
where $z^{l(s)}_{max}$ is the maximum amplitude of the LS (SS) and $Q_{tr}(\omega)$ is the fluctuational noise-induced spectral peak that arises due to the noise-induced transition between the two dynamical states.
Figure~\ref{Ch5_Fig2}(b) clearly shows the maximum SDF intensity near the driving frequency 35 Hz, as expected.
\subsection{Interatomic interaction as one-way bias field}
\begin{figure}[ht]\centering
\includegraphics[scale=0.5]{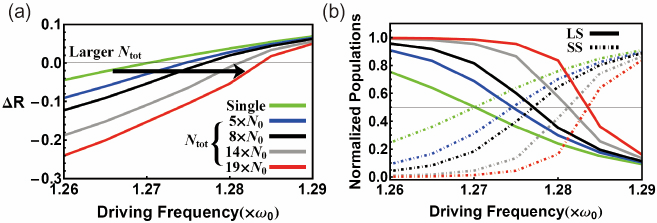}
\caption[figure23]{\label{Ch5_Fig3}\small Simulation results for the activation energy and the atomic population versus the driving frequency. (a) The activation energy difference $\Delta R$ between the LS and the SS decreases with the interaction or $N_{\mathrm{tot}}$. (b) The change of the normalized  populations in the LS and the SS  in the weakly interacting regime. Here $N_0=10^6$, and $\gamma$ and $\omega_{0}$ are 40.58 s$^{-1}$ and $2\pi\times32.68$ Hz, respectively. (Figures from Ref.~\cite{gmoon njp2013})}
\end{figure}

As mentioned in Sec.~\ref{Switching rate modification}, the atom-atom interaction produces the modification of the activation energy, which is measurable because the cumulative effect of the inter-particle interaction between the two clouds becomes significant~\cite{JETF 1979, M Heo PRE 2010}.
Figure \ref{Ch5_Fig3} presents the theoretical results of the activation energy (Eq.~(\ref{activationdiff})) and the associated atomic populations (Eq.~(\ref{population1})) due to the atom-atom interaction.
As discussed in Sec.~\ref{sec 3.2}, the long-range attractive shadow force increases with $N_{\mathrm{tot}}$  and 
as previously shown in Fig.~\ref{Ch5_Fig3} (a), the activation energy of the LS (SS) increases (decreases) with  $N_{\mathrm{tot}}$. Furthermore, one obviously observes  that the change of $R_{1(2)}$ shifts the KPT point, where $\Delta R~(=R_{2}-R_{1})=0$, toward the higher $\omega_F$ at the higher  $N_{\mathrm{tot}}$. In addition, at the fixed modulation frequency $\omega_F$, $\Delta R$ decreases with the increase of $N_{\mathrm{tot}}$ and thereby the resulting population changes as shown in Fig.~\ref{Ch5_Fig3}(b); $\mathcal{P}_1$ of the LS increases with  $N_{\mathrm{tot}}$ (solid line) whereas $\mathcal{P}_2$ of the SS decreases (dot-dashed line), shifting the KPT point toward the higher $\omega_F$.
\begin{figure}[ht]\centering
\includegraphics[scale=0.45]{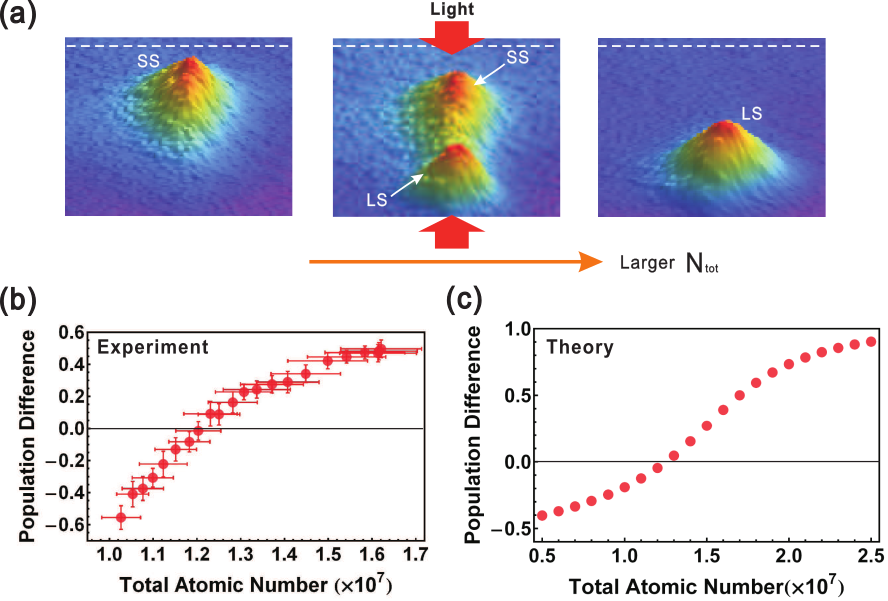}
\caption[figure24]{\label{Ch5_Fig4}\small The change of the atomic-population difference resulting from the atom-atom interaction between two stable states at the driving frequency 34 Hz. (a) presents the atomic images displaying the change of atomic population with $N_{\mathrm{tot}}$. (b) and (c) present the experimental  and theoretical results for the population difference of atoms, $\mathcal{P}_1-\mathcal{P}_2$, between the two coexisting dynamical states, respectively. The white dashed line indicates the trap center. (Figures from Ref.~\cite{gmoon njp2013})}
\end{figure}

We have experimentally observed the effects of the atom-atom interactions on the single-particle behavior of the KPT (i.e. KPT without interaction). As shown in Fig.~\ref{Ch5_Fig4}(a), when  $N_{\mathrm{tot}}$ is increased, atoms in the SS are transferred to the LS in a unidirectional way. This is because the interaction plays the role as the one-way bias field due to the asymmetry of the noise-induced optimal path of two stable states, in contrast to the effect of the interaction in the PDDO that induces the time-translational symmetry breaking, which then results in the ideal mean-field transition.
The difference of the atomic populations $\mathcal{P}_1-\mathcal{P}_2$ in Fig.~\ref{Ch5_Fig4}(b) demonstrates the effect of the interaction in the KPT as the one-way bias field, displaying that the population difference grows larger at the higher $N_{\mathrm{tot}}$. The experimental results are in qualitatively good agreement with the theoretical results in Fig.~\ref{Ch5_Fig4}(c).

\subsection{Phase-transition diagram in the parameter space}
\begin{figure}[ht]\centering
\includegraphics[scale=0.5]{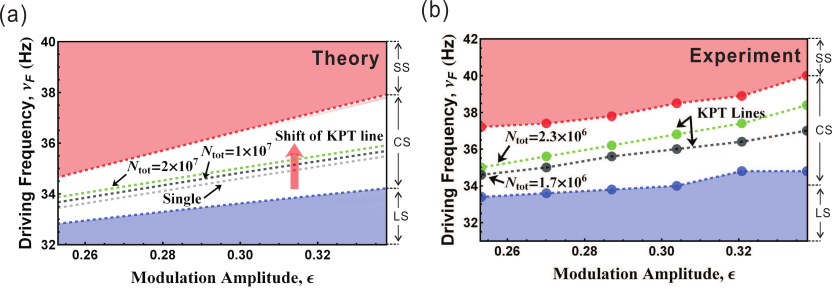}
\caption[figure25]{\label{Ch5_Fig5}\small (a) Theoretical results for the shift of the KPT phase boundary in the $\nu_F$-$\epsilon$ parameter space with respect to the cumulative effect of the atom-atom interaction (or $N_{\mathrm{tot}}$). (b) Experimental results for the shift of the KPT boundary.
The measured $\gamma$ and $\omega_{0}$ are 40.56 s$^{-1}$ and $2\pi\times28.07$ Hz, respectively. The red and blue dashed lines represent the boundary where the transition between the monostable state (LS or SS) and the coexisiting bistable state (CS) occurs. Note that these boundary lines remain unaffected under the weak atom-atom interaction. As shown in the CS region, the typical KPT lines shift from the black to  green line as $N_{\mathrm{tot}}$ is increased. Each circle denotes the experimental data, which show qualitatively good agreement with the numerical results except the differing absolute values of $N_{\mathrm{tot}}$. (Figures from Ref.~\cite{gmoon njp2013})}
\end{figure}

One can draw the phase-transition map in the parameter space of the driving frequency $\omega_F$ and the modulation amplitude $\epsilon$. In Fig.~\ref{Ch5_Fig5}(a), the calculated KPT lines (black and green dashed lines) are shown, where the changes of the KPT boundary are expected to depend on the total number of trapped atoms. The experimental results in Fig.~\ref{Ch5_Fig5}(b) are in qualitatively good agreement.

Interestingly, the similar phenomenon has been observed in the discontinuous gas-liquid phase transition in equilibrium states~\cite{Kunz,Goldenfeld}. The gas-liquid transition described by the van der Waals equation takes into account the nonzero radius of  atoms as well as the attractive interactions between atoms. In the pressure-temperature ($P$-$T$) map, the phase transition can be plotted where the coexistence region and the critical point can be identified. The atomic radius and the interaction depend on the species of atom, and thus the coexistence line in the $P$-$T$ plane changes with respect to the atomic species. Therefore, the attractive interaction between the  RDDO atoms in the coexisting periodic attractors out of equilibrium exhibits a very close similarity with the gas-liquid transition in equilibrium. 

\section{Summary and outlook}
We have reviewed the collective and the nonequilibrium phenomena for the various nonlinear dynamical systems, the PDDO and the RDDO realized in the cold $^{85}$Rb MOT  system. The cold atoms (T $\sim$ a few hundreds $\mu$K) congregate around the stable state (attractor), and they undergo the noise-induced switching between the two dynamical stable states by thermal fluctuations. Interestingly there exists the globally coupled attractive interaction among atoms, which induces the symmetry-breaking shadow force. The interatomic interaction and the large rare fluctuation provides the numerous qualitative and quantitative characteristics of the nonlinear dynamical system out of equilibrium.

In Sec. 2, we have analyzed the nonlinear dynamics using the cold atom trap modulated in various ways, and we have also compared a several methods for the trap-parameter measurement; the transient oscillation method, the parametric resonance method and the forced harmonic oscillation method. In Sec. 3, we have discussed the noise-induced switching dynamics of many particles with the inter-particle interaction considered, and derived the master equation displaying the time-translational symmetry breaking and the role of the inter-particle interaction in the kinetic phase transition.

In Sec. 4, for the PDDO with the global coupling interaction, we have described the novel property of the time-translational symmetry breaking in terms of the ideal mean-field model, and we have confirmed this analogy by the critical exponent measurements. Furthermore, in the PDDO we have experimentally and theoretically discussed the scaling behavior of the relaxation process for an unstable state near the subcritical Hopf bifurcation point. Near such a bifurcation point, we have experimentally shown that the relaxation process exhibits the scaling behavior; the relaxation time shows the scaling exponent of $-1.002$ ($\pm$0.024). In addition, we have  discussed the hysteresis of the spontaneous symmetry breaking transition obtained by sweeping the total number of atoms and dealt with the thermal hysteretic behavior by the scaling exponent measurement of the hysteresis. It is also shown that the relaxation time of the order parameter becomes longer near the critical point. The scaling exponent of the hysteresis area with the atomic number sweeping rate is found to be $0.64 \pm 0.04$, which is consistent with the value derived by the mean-field model.

In Sec. 5, in the RDDO we have presented the experimental work on the kinetic phase transition realized in the many-body system, which is manifested by the substantial enhancement of fluctuations. Such an enhancement results from the noise-induced switching between the coexisting states, similarly to the first-order phase transition in the thermal equilibrium system. Moreover, we could control the attractive interaction between the atoms trapped in the bistable states by adjusting the total number of atoms, which serves as the one-way bias field that produces the unilateral transfer of atoms. The interatomic interaction thereby has induced the shift of the phase-transition boundary. These results provide another piece of evidence for the similarity of phase transitions between equilibrium and nonequilibrium systems.

It is worth while to emphasize that the nonequilibrium system undergoes an ideal mean-field phase transition  due to the interplay of the noise and the nonlinearity associated with the interparticle interaction.
In particular, the results demonstrate that the notion of discrete symmetry breaking transitions can be extended to the time domain, where the time-translation symmetry breaking occurs due to the modification of switching dynamics by the interatomic interaction. Such a symmetry breaking feature may be further investigated in a close relation to the recent works on the time crystal; classical vs quantum and discrete vs continuous.
The results summarized in this review may be  also helpful for better understanding of the fundamental physics of critical phenomena occurring in a many-body system far from thermal equilibrium, which is an interesting and important subject  to be addressed further both experimentally and theoretically.   

\section*{Acknowledgments}
This work was supported by the National Research Foundation of Korea (NRF) grant funded by the Korea government (MSIP) (No. 2016R1A3B1908660).
The authors are grateful to Prof. Kihwan Kim who had performed the initial series of experiments on the parametrically driven cold atoms including the direct observation of the spontaneous symmetry breaking  therein.

\end{document}